\documentclass[apjl]{emulateapj}
\usepackage{amsfonts,amsmath,graphicx,natbib,apjfonts,subfigure,textcomp,color}

\def\nu{1}
\def\nyu{2}
\def\col{3}
\def\ou{4}
\def\cfa{5}
\def\fer{6}
\def\fr{7}
\def\cal{8}
\def\deanne{9}
\def\sb{10}
\def\maria{11}
\def\psu{12}
\def\inafmi{13}

\shorttitle{The bactrian transient ASASSN-15lh}
\shortauthors{Margutti et al.}

\begin{document}
%\title{X-rays, variability and an active galaxy at the location of the Bactrian Transient ASASSN-15lh}
\title{X-rays from the location of the Bactrian Transient ASASSN-15lh}%: \\ Implications for an unprecedented Tidal Disruption}

\author{R. Margutti\altaffilmark{\nu,\nyu}, 
B.~D. Metzger\altaffilmark{\col}, 
R. Chornock\altaffilmark{\ou},  
D. Milisavljevic\altaffilmark{\cfa}, 
E. Berger\altaffilmark{\cfa}, 
P. K. Blanchard\altaffilmark{\cfa}, 
C. Guidorzi\altaffilmark{\fer}, 
G. Migliori\altaffilmark{\fr},
A. Kamble\altaffilmark{\cfa},
R. Lunnan\altaffilmark{\cal},
M. Nicholl\altaffilmark{\cfa}, 
D.~L. Coppejans\altaffilmark{\deanne},
S. Dall'Osso\altaffilmark{\sb},
M.~R. Drout\altaffilmark{\maria,*}, 
R. Perna\altaffilmark{\sb},
B. Sbarufatti\altaffilmark{\psu,\inafmi}
}

\altaffiltext{\nu}{Center for Interdisciplinary Exploration and Research in Astrophysics (CIERA) and Department of Physics and Astronomy, Northwestern University, Evanston, IL 60208}
\altaffiltext{\nyu}{Center for Cosmology and Particle Physics, New York University, 4 Washington Place, New York, NY 10003, USA}
\altaffiltext{\col}{Columbia Astrophysics Laboratory, Columbia University, Pupin Hall, New York, NY 10027, USA}
\altaffiltext{\ou}{Astrophysical Institute, Department of Physics and Astronomy, 251B Clippinger Lab, Ohio University, Athens, OH 45701, USA}
\altaffiltext{\cfa}{Harvard-Smithsonian Center for Astrophysics, 60 Garden St., Cambridge, MA 02138, USA}
\altaffiltext{\fer}{Department of Physics and Earth Sciences, University of Ferrara, via Saragat 1, I-44122, Ferrara, Italy}
\altaffiltext{\cal}{Department of Astronomy, California Institute of Technology, 1200 East California Boulevard, Pasadena, CA 91125, USA}
\altaffiltext{\fr}{Laboratoire  AIM  (CEA/IRFU  -  CNRS/INSU  -  Universite  Paris  Diderot),  CEA  DSM/SAp,  F-91191  Gif-sur-Yvette,
France}
\altaffiltext{\deanne}{Department of Astrophysics/IMAPP, Radboud University, P.O. Box 9010, 6500 GL Nijmegen, The Netherlands}
\altaffiltext{\maria}{Carnegie Observatories, 813 Santa Barbara Street, Pasadena, CA 91101, USA 2}
\altaffiltext{\sb}{Department of Physics and Astronomy, Stony Brook University, Stony Brook, NY, 11794, USA}
\altaffiltext{\psu}{Department of Astronomy and Astrophysics, Pennsylvania State University, 525 Davey Lab, University Park, PA 16802, USA}
\altaffiltext{\inafmi}{INAF, Osservatorio Astronomico di Brera, via E. Bianchi 46, 23807 Merate, Italy}
\altaffiltext{*}{Hubble, Carnegie-Dunlap Fellow}

\begin{abstract}
We present the detection of persistent soft X-ray radiation with $L_x\sim10^{41}-10^{42}\,\rm{erg\,s^{-1}}$ at the location of the extremely luminous, double-humped transient ASASSN-15lh as revealed by \emph{Chandra}  and \emph{Swift}. We interpret this finding in the context of observations from our multiwavelength campaign, which revealed the presence of weak narrow nebular emission features from the host-galaxy nucleus and clear differences with respect to superluminous supernova optical spectra.
Significant UV flux variability on short time-scales detected at the time of the re-brightening disfavors the shock interaction scenario as the source of energy powering the long-lived UV emission, while deep radio limits exclude the presence of relativistic jets propagating into a low-density environment.
We propose a model where the extreme luminosity and double-peaked temporal structure of ASASSN-15lh is powered by a central source of ionizing radiation that produces a sudden change of the ejecta opacity at later times. As a result, UV radiation can more easily escape, producing the second bump in the light-curve.  We discuss different interpretations for the intrinsic nature of the ionizing source.% that powers ASASSN-15lh.
%, either a stellar explosion with extreme properties of the disruption of a star by the  host-galaxy supermassive black hole.  
We conclude that, \emph{if} the X-ray source is physically associated with the optical-UV transient, ASASSN-15lh most likely represents the tidal disruption of a main-sequence star by the most massive spinning black hole detected to date. 
%the first detection of emission from the  tidal disruption of a main sequence star by a very massive, spinning supermassive black hole. 
In this case, ASASSN-15lh and similar events discovered in the future would constitute the most direct probes 
%the massive end of dormant, spinning, supermassive black holes in galaxies. \\
of very massive, dormant, spinning, supermassive black holes in galaxies. Future monitoring of the X-rays may allow us to distinguish between the
supernova and TDE hypothesis. \\
\end{abstract}
%%%%%%%%%%%%%%%%%%%%%%%%%%%%%%%%%%%%%%%%%%%
\section{Introduction}
Optical surveys sampling the sky over time scales of a few days significantly advanced our knowledge of astronomical transients of different origins,
%the transient sky and recently led to the discovery of new classes of explosive transients, 
including super-luminous supernovae (SLSNe; \citealt{Quimby11, Chomiuk11, Gal-Yam12}), very fast-rising stellar explosions (e.g. \citealt{Drout14}) and stellar tidal disruptions by super-massive black holes (TDEs; \citealt{Rees88,Komossa15}). Occasionally, a transient is found with properties that seem to defy all previous classification schemes. The event ASASSN-15lh belongs to this category.

ASASSN-15lh  (\citealt{Dong16}) was discovered by the All-Sky Automated Survey for Supernovae (ASAS-SN\footnote{http://www.astronomy.ohio-state.edu/~assassin/index.shtml}) on 14 June 2015 at z=0.2326  ($d=1171$ Mpc for standard \emph{Plank} cosmology). Its extremely large peak luminosity $L_{pk}\sim2\times10^{45}\,\rm{erg\,s^{-1}}$  and the blue, almost featureless spectrum with no apparent sign of H or He (and some spectroscopic resemblance to the SLSN 2010gx) led \cite{Dong16} and \cite{Godoy-Rivera16} to suggest that ASASSN-15lh is the most luminous SLSN ever detected. The very large energy radiated by ASSASN-15lh ($E_{rad}\sim (1.5-2)\times10^{52}\,\rm{erg}$, \citealt{Godoy-Rivera16}), requires extreme properties of the progenitor star  and sources of energy that are different from the standard radioactive decay of $^{56}$Ni that powers normal H-stripped SNe in the local Universe (\citealt{Dong16, Chatzopoulos16,Kozyreva16}). In this context, the double-humped light-curve of ASASSN-15lh has been interpreted by \cite{Chatzopoulos16} as a signature of the interaction of massive SN ejecta $M_{ej}\sim36\,\rm{M_{\sun}}$ with an H-poor circumstellar shell of $M_{CSM}\sim20\,\rm{M_{\sun}}$, possibly supplemented by radiation from a newly-born rapidly-rotating magnetar \citep{Metzger15,Sukhbold16,Bersten16,Dai+16}.

The old, massive $M_{\ast}\sim 2\times 10^{11}\,\rm{M_{\sun}}$ host galaxy of ASASSN-15lh, with limited star formation ($SFR<0.3\,\rm{M_{\sun}yr^{-1}}$; \citealt{Dong16}), is however markedly different from those of core-collapse SNe (e.g. \citealt{Leaman11}) as well as of envelope-stripped SLSNe, which tend to be younger star forming systems with significantly lower stellar mass (\citealt{Lunnan14,Lunnan15,Perley16}). This observation, together with the location of the transient, astrometrically consistent with the host galaxy nucleus, inspired a connection between ASASSN-15lh and the tidal disruption of a star by the host-galaxy super-massive black hole (SMBH, \citealt{Godoy-Rivera16,Brown16,leloudas16,Perley16}). In this context ASASSN-15lh would be the most luminous TDE ever observed, associated to a SMBH with mass $M_{\bullet}\sim10^{8.6}\,\rm{M_{\sun}}$ \citep{Dong16, Godoy-Rivera16} significantly larger than any SMBH presently associated to a TDE (e.g. \citealt{Komossa15}).

It is clear that the luminosity, the spectral properties and the double-humped (i.e., ``Bactrian'') light-curve of ASASSN-15lh, as well as its host galaxy, are unprecedented both in the context of SLSNe and in the context of TDEs. 

In this paper we present and discuss the following observational facts: [i] Uncovering of persistent, soft X-ray emission at the location of ASASSN-15lh (Sec. \ref{SubSec:CXO} and Sec. \ref{SubSec:XRT}); [ii] Detection of significant temporal variability at UV wavelengths during the re-brightening phase (Sec. \ref{SubSec:UV}); [iii] Detection of narrow nebular spectral features connected to the host-galaxy nucleus (Sec. \ref{SubSec:Spec}). We propose a scenario where a single physical mechanism can naturally explain the double-humped light-curve of ASASSN-15lh and suggest that its location, very close or coincident with the nucleus of a galaxy that harbors a massive SMBH, is likely the key to unlocking the mysterious nature of the transient (Sec. \ref{Sec:Int}). 
Conclusions are drawn in Sec. \ref{Sec:conclusions}.

In our analysis we assume the object's time of first light to be  April 29, 2015, corresponding to a 30-day (rest-frame) rise-time to maximum V-band luminosity \citep{Dong16}. Our main conclusions do not depend on this assumption.
%MJD=57141.5 Explosion date --> April 29th, 2015
%%%%%%%%%%%%%%%%%%%%%%%%%%%%%%%%%%%%%%%%%%%
\section{Data Analysis and Results}

%%%%%%%%%%%%%%%%%%
\subsection{X-ray analysis- CXO}
\label{SubSec:CXO}

\begin{figure}
\vskip -0.0 true cm
\centering
\includegraphics[scale=0.42]{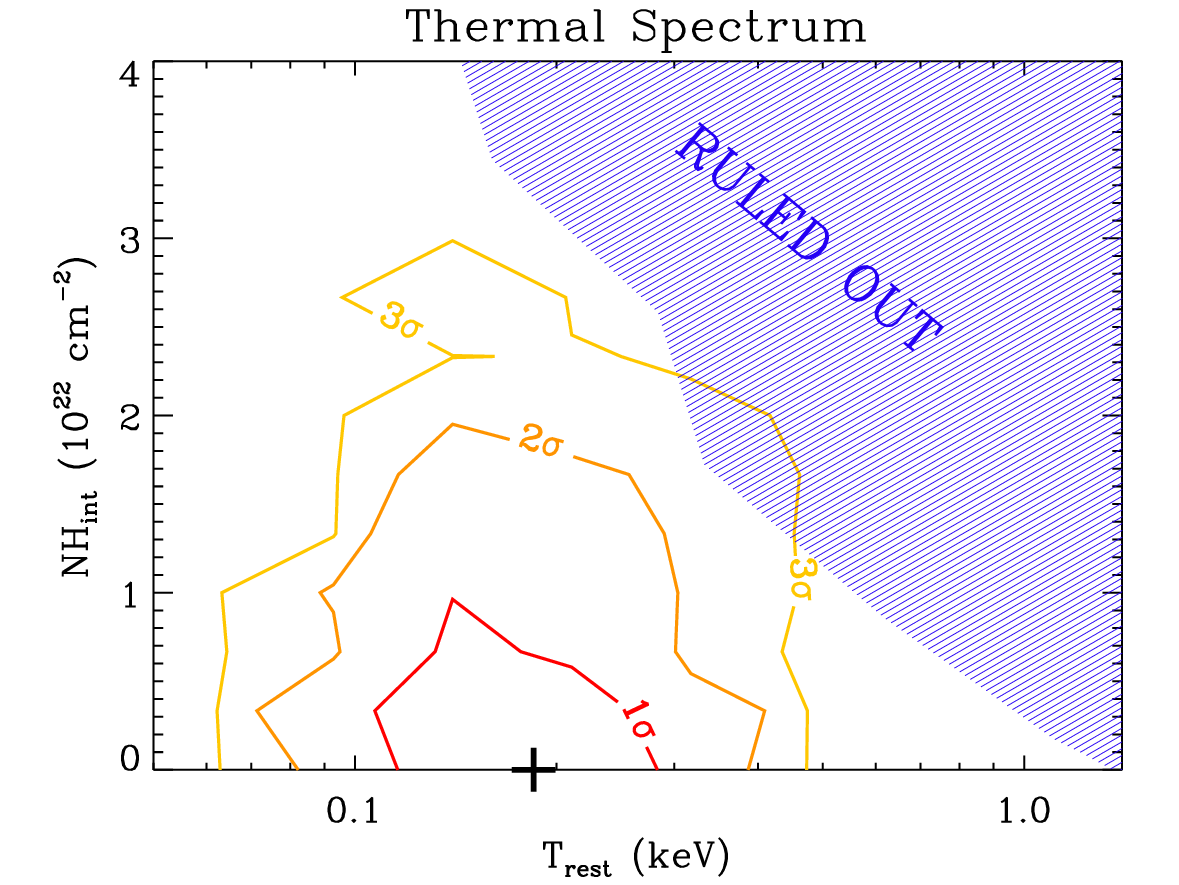}
\includegraphics[scale=0.42]{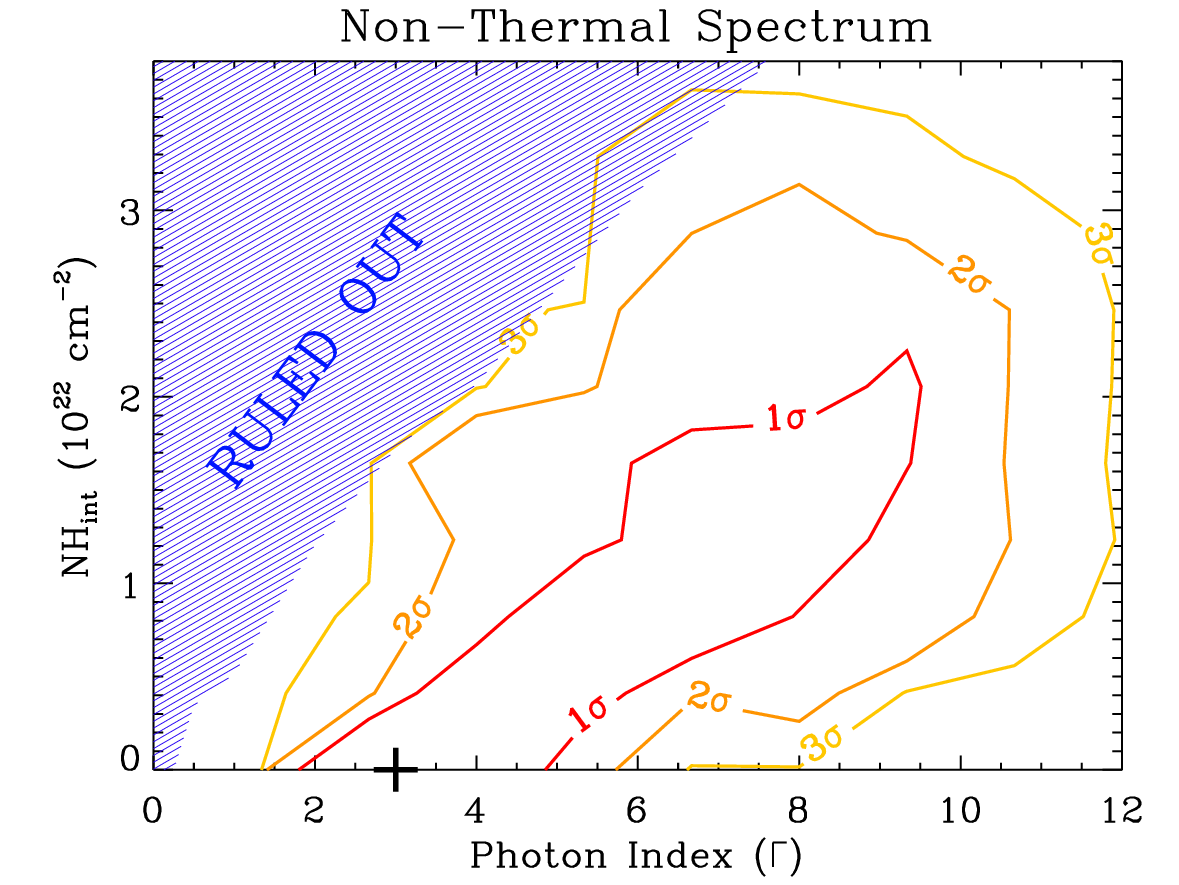}
\caption{Constraints on the spectral model parameters of the X-ray source detected at the location of ASASSN-15lh, as derived from our second epoch (shaded region) and a joint fit of the third and fourth epoch (contours) of CXO observation. \emph{Upper panel}: absorbed blackbody spectrum. \emph{Lower panel}: absorbed power-law spectrum. In both panels the cross symbol identifies the best fitting model parameters. These observations favor a soft X-ray spectrum with limited intrinsic absorption. }
\label{Fig:Chandra}
\end{figure}

We obtained four epochs of deep X-ray observations of ASASSN-15lh with the Chandra X-ray Observatory (CXO) on November 12, 2015 (exposure of 10 ks), December 13, 2015 (10 ks), February 20, 2016 (40 ks) and August 19, 2016 (30 ks, PI Margutti), corresponding to $\delta t$$=$$129.4$ days, $\delta t$$=$$154.6$ days, $\delta t$$=$$210.5$ and $\delta t$$=$$357.8$ days rest frame since optical maximum light, which occurred on June 5, 2015 \citep{Dong16}. CXO data have been reduced with the CIAO software package (version 4.8) and corresponding calibration files. Standard ACIS data filtering has been applied. 

ASASSN-15lh is not detected in our first epoch of observations (ID 17879),  with a 3$\sigma$ count-rate upper limit of $9.98\times 10^{-5}\,\rm{c\,s^{-1}}$ (0.5-8 keV).  The Galactic column density in the direction of the transient is $3.07\times 10^{20}\,\rm{cm^{-2}}$ \citep{Kalberla05}. For an assumed power-law spectrum with  photon index $\Gamma=2$ and Galactic absorption, the unabsorbed 0.3-10 keV flux limit is $F_x<1.1\times 10^{-15}\rm{erg\,s^{-1}cm^{-2}}$ ($L_x<1.8\times 10^{41}\,\rm{erg\,s^{-1}}$). % For a blackbody spectrum with $T=1$ keV and Galactic absorption, the unabsorbed 0.3-10 keV flux limit is $F_x<1.2\times 10^{-15}\rm{erg\,s^{-1}cm^{-2}}$ ($L_x<2.0\times 10^{41}\,\rm{erg\,s^{-1}}$). 
Our analysis below favors a soft X-ray spectrum with negligible absorption and $\Gamma\sim 3$ or a thermal spectrum with $T\sim0.17\,\rm{keV}$. For these parameters, the unabsorbed 0.3-10 keV flux limit is $F_x<2.0\times 10^{-15}\rm{erg\,s^{-1}cm^{-2}}$ (non-thermal spectrum) and $F_x<8.0\times 10^{-16}\rm{erg\,s^{-1}cm^{-2}}$ (thermal spectrum).

In our second epoch of observation (ID 17880) we find evidence for weak, soft X-ray emission at the location of ASASSN-15lh. We detect two photons with energy $<$1 keV in a $1\arcsec$ region around the transient, corresponding to a 4.3$\,\sigma$ c.l.  detection in the 0.5-1 keV energy range, and to a 3.3$\,\sigma$ c.l. detection in the 0.5-8 keV energy range. We constrain the spectral parameters by using the observed background and the actual instrumental response to simulate the expected emission from a grid of thermal and non-thermal spectral models with a wide range of intrinsic absorption  $NH_{int}=(0-4)\times 10^{22}\,\rm{cm^{-2}}$. The regions excluded at 3$\sigma$ confidence are shaded in Fig. \ref{Fig:Chandra}.

An X-ray source is clearly detected at the location of ASASSN-15lh at the time of our third CXO observation (ID 17881), with count-rate $1.5\times 10^{-4}\,\rm{c\,s^{-1}}$ and significance of 5.2$\sigma$ in the 0.5-8 keV range (4.7$\sigma$ in the 0.5-1 keV energy range). In our fourth epoch (ID 17882) the source is still detected with count-rate $1.7\times 10^{-4}\,\rm{c\,s^{-1}}$ and significance of 4.9$\sigma$ in the 0.5-8 keV range (3.6$\sigma$ in the 0.5-1 keV energy range). 

We employ the Cash statistics to fit the spectra (we have a total of six and five photons in a 1$\arcsec$ region around the transient in the third and fourth epoch, respectively), and perform a series of MCMC simulations to constrain the spectral parameters. The analysis of each of the two epochs taken separately points to a soft X-ray spectrum, with limited absorption and no evidence for statistically significant evolution between the two epochs. We thus constrain the X-ray source spectral parameters with a joint spectral fit of the two CXO epochs of observation, where the spectral normalization is allowed to vary from one epoch to the other. 

The results are displayed in Fig. \ref{Fig:Chandra}. For an absorbed,  non-thermal power-law spectrum, the best-fitting parameters are $\Gamma=3.0$ and $NH_{int}\sim 10^{19}\,\rm{cm^{-2}}$. The inferred (0.3-10 keV) unabsorbed flux for this model is $F_x=3.6\times 10^{-15}\rm{erg\,s^{-1}cm^{-2}}$ corresponding to $L_x=5.9\times 10^{41}\,\rm{erg\,s^{-1}}$ (third epoch) and $F_x=4.9\times 10^{-15}\rm{erg\,s^{-1}cm^{-2}}$ ($L_x=8.1\times 10^{41}\,\rm{erg\,s^{-1}}$, fourth epoch).
The best-fitting parameters for an absorbed blackbody spectrum are: $T=0.17$ keV, $NH_{int}\sim 10^{18}\,\rm{cm^{-2}}$. The inferred (0.3-10 keV) unabsorbed flux for this model is $F_x=1.2\times 10^{-15}\rm{erg\,s^{-1}cm^{-2}}$, corresponding to $L_x=2.0\times 10^{41}\,\rm{erg\,s^{-1}}$ (third epoch) and $F_x=1.4 \times 10^{-15}\rm{erg\,s^{-1}cm^{-2}}$ ($L_x=2.3\times 10^{41}\,\rm{erg\,s^{-1}}$, fourth epoch). Both spectral models point to a very limited amount of neutral Hydrogen in the host galaxy along our line of sight, consistent with the very low N(HI) inferred by \cite{leloudas16} from Ly-$\rm{\alpha}$ and the very strong high ionization lines (\ion{N}{5} and \ion{O}{6}).

With reference to Fig. \ref{Fig:Chandra} we find that: (i) the X-ray source shows a soft spectrum (most of the allowed parameter space is  at $\Gamma>2$ and $T<1$ keV) with limited intrinsic absorption (of the order of a few $10^{22}\,\rm{cm^{-2}}$ at most).  (ii) There is no evidence for strong temporal and/or spectral variability of the X-ray source. 

We first evaluate the possibility that the X-ray emission arises from a population of Low-Mass X-ray Binaries (LMXBs) residing in the early-type host galaxy, using the $L_x-L_{\rm{B}}$ and $L_x-L_{\rm{K}}$ relations by  \cite{Kim04}. For the host-galaxy of ASASSN-15lh \cite{Dong16} measure $M_{\rm{K}}=-25.5$ mag and $M_{\rm{B}}=-19.96$ mag, which imply $L_{x,XRB}=(1-6)\times 10^{40}\,\rm{erg\,s^{-1}}$ (0.3-8 keV). This is a factor $\ge10$ smaller then the measured X-ray emission at the location of ASASSN-15lh (re-calibrated with the same spectral model as \citealt{Kim04} in the 0.3-8 keV band). We conclude that LMXBs are unlikely to be the source of the detected X-rays. We thus envision two possible scenarios: either the X-rays originate from weak AGN activity from the host galaxy nucleus or they are physically connected to the optical/UV transient. In the first case we expect a somewhat stable X-ray emission over the time scale of years, while we anticipate fading if the X-ray emission is directly connected to ASASSN-15lh. Future observations will clarify the origin of the detected high-energy emission. Below we put our results into the context of X-ray emission from known transients (i.e. SNe and TDEs).

\begin{figure*}
\vskip -0.0 true cm
\centering
\includegraphics[scale=0.7]{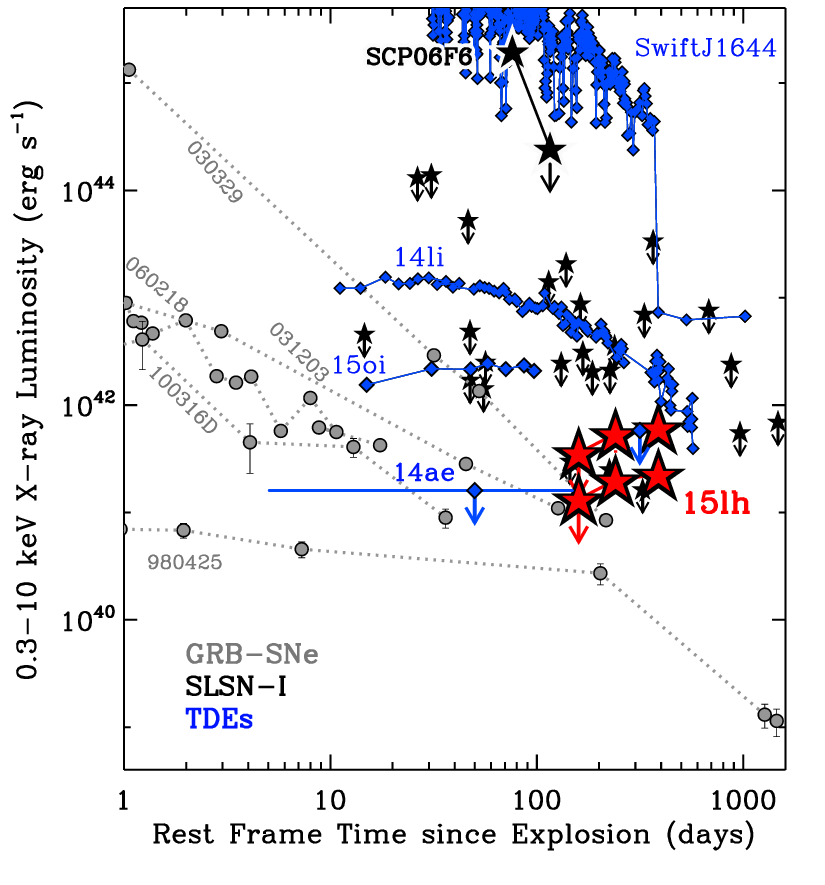}
\caption{ASASSN-15lh in the X-ray phase space of envelope-stripped SNe and TDEs. For ASASSN-15lh we merged the second and third CXO epoch for the sake of clarity and we show the luminosity of ASASSN-15lh at each epoch for both a thermal (bottom points) and a non-thermal (top points) spectrum. At $\sim 100$ days, ASASSN-15lh is more luminous than any previously detected H-stripped SN with the exception of SCP06F6. Its persistent and soft X-ray emission is more similar to non-relativistic TDEs, like ASASSN-15oi and ASASSN-14li.
The X-ray light-curve of the relativistic TDE SwiftJ2058 overlaps with the TDE SwiftJ1644 \citep{Pasham15} and it is not displayed here for clarity.
References: \cite{Gezari12,Margutti13d,Holoien14,Miller15,Holoien16,Mangano16,Brown16bis}. For SLSNe-I  we updated the sample of \cite{Levan13}. The detailed analysis will appear in Margutti et al., in prep.}
\label{Fig:XraysLC}
\end{figure*}

The detected emission is softer than the typical X-ray spectrum of SNe associated with Gamma-Ray Bursts ($\Gamma\sim2$, e.g. \citealt{Margutti13}) and normal H-stripped SNe (e.g. \citealt{Chevalier06,Dwarkadas12}), which typically show $\Gamma\sim2$ and a decaying flux with time. A way to sustain luminous X-ray emission over a long time is to invoke the SN shock interaction with a thick medium (see e.g. SN\,2014C, \citealt{Margutti16}). However, the observed X-ray spectrum of H-stripped SNe strongly interacting with the environment is even harder ($T\sim20$ keV), and thus even more different from what we observe at the location of ASASSN-15lh  (e.g. \citealt{Margutti16}). It is thus unlikely that a SN shock interaction with the medium is powering both the X-ray and optical/UV emission from ASASSN-15lh. Finally, compared to the only other X-ray source associated to a SLSN-I so far, the emission at the location of ASASSN-15lh is also  softer and significantly longer lived (Fig. \ref{Fig:XraysLC}): for the SLSN-I SCP06F6, \cite{Levan13} reports $\Gamma\sim2.6 $ (or a thermal spectrum with $T\sim1.6$ keV). 

The X-ray properties of ASASSN-15lh are instead more reminiscent of the soft X-ray emission detected in non-jetted TDEs.  Non-jetted TDEs detected with \emph{ROSAT}, \emph{XMM-Newton}, \emph{Chandra} and, more recently, with \emph{Swift} show peak luminosities of $L_x\sim10^{42}-10^{44}\,\rm{erg\,s^{-1}}$ and very soft spectra that later harden with time on a time-scale of years and with initial temperatures $T<0.2$ keV  (e.g. \citealt{Komossa15} for a recent review).

As for the TDEs ASASSN-15oi \citep{Holoien16} and ASASSN-14li \citep{vanVelzen16}, the X-ray emission is more luminous than what expected based on the extrapolation of the optical/UV blackbody model (see Sec. \ref{SubSec:UV}) and a more complex model is needed. In this context ASASSN-15lh would show the most extreme ratio $L_{\nu,UV}/L_{\nu,X-rays}\sim 10^5$ (compared to $L_{\nu,UV}/L_{\nu,X-rays}\sim 10^{2-3}$ for ASASSN-14li and $L_{\nu,UV}/L_{\nu,X-rays}\sim 10^4$ for ASASSN-15oi).

\begin{figure}
\vskip -0.0 true cm
\centering
\includegraphics[scale=0.6]{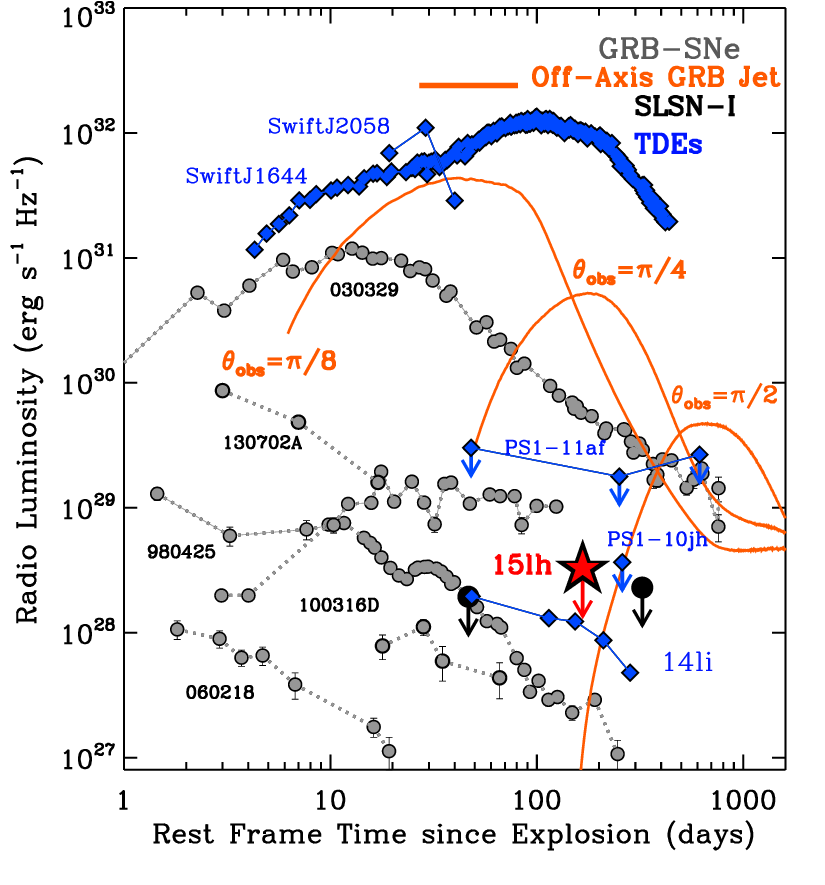}
\caption{ASASSN-15lh (red star) in the context of radio emission from envelope-stripped SNe (grey and black filled circles for GRB-SNe and SLSNe, respectively) and TDEs (blue diamonds), as constrained by observations with ATCA \citep{Kool15}. Radio observations acquired by \citet{leloudas16} three weeks later led to very similar limits and are not displayed here for clarity. The limits on the radio emission from ASASSN-15lh rule out most of the parameter space associated with powerful relativistic jets with kinetic energy $E_k$=$10^{51}\,\rm{erg}$ propagating into a circumburst medium with density  $1\,\rm{cm^{-3}}$ (orange lines, \citealt{vanEerten10}). SwiftJ1644 and SwiftJ2058 are the two relativistic TDEs known to date with radio observations.  References: \cite{Cenko12,Chomiuk12b,Chandra12,vanVelzen13,Margutti13d,Chornock14,Nicholl16,Alexander16,Berger12,Pasham15,Zauderer13}. }
\label{Fig:Radio}
\end{figure}

In Fig. \ref{Fig:XraysLC} we put ASASSN-15lh on the X-ray luminosity plane of energetic envelope-stripped core-collapse SNe (i.e. GRB-SNe and SLSNe) and TDEs.  ASASSN-15lh is $\sim1000$ times less luminous than the SLSN-I SCP06F6  and does not experience a similar drop in luminosity. At $\sim$100 days, the X-ray emission at the location of ASASSN-15lh is more luminous than GRB-SNe. However, observations obtained around the same epoch by the ATCA in Fig.  \ref{Fig:Radio} put deep limits to the radio emission from ASASSN-15lh \citep{Kool15,leloudas16}, and rule out the presence of powerful jets seen on axis (most of the parameter space associated with off-axis GRB-like jets is also ruled out).  Also in this case, the luminous and not strongly variable X-ray emission at the location of ASASSN-15lh, which lacks a luminous radio counterpart, seems to be more in line with observations of non-jetted TDEs (recent examples are ASASSN-14li, \citealt{Miller15, Alexander16,Holoien16}; or ASASSN-15oi, \citealt{Holoien16b}).

%----------------
\subsection{X-ray analysis- Swift/XRT and XMM-Newton}
\label{SubSec:XRT}

We re-processed all the X-ray data collected by the \emph{Swift}-XRT \citep{Burrows05} between June 24, 2015 until July 22, 2016 (total exposure time of $\sim$270 ks), following the prescriptions outlined in \cite{Margutti13}. A targeted search for X-ray emission at the location of ASASSN-15lh identifies the presence of a weak X-ray excess with significance of $3\,\sigma$ in the 0.3-5 keV range. The significance is reduced to $2.4\,\sigma$ in the 0.3-10 keV energy range, consistent with the soft X-ray spectrum suggested by the CXO observations. We infer a background subtracted count-rate of $(1.1\pm0.4)\times 10^{-4}\,\rm{c\,s^{-1}}$ (0.3-5 keV), which corresponds to an unabsorbed 0.3-10 keV flux $F_x=(4.1\pm1.5)\times 10^{-15}\,\rm{erg\,s^{-1}cm^{-2}}$ and $F_x=(3.7\pm 1.4)\times 10^{-15}\,\rm{erg\,s^{-1}cm^{-2}}$ for a blackbody and power-law spectral model, respectively, and the best-fitting spectral parameters derived from the CXO data. The average flux inferred from \emph{Swift}-XRT observations is thus consistent with the results from the CXO analysis and suggests that the X-ray source at the location of ASASSN-15lh experienced at most mild temporal variability over the $\sim$1 yr of \emph{Swift} monitoring. We note that flux variations of the order of a factor of a few are consistent with our findings, given the uncertainties affecting both the  \emph{Swift}-XRT and the CXO measurements. A delayed onset of the X-ray emission with respect to the optical emission is also clearly allowed, since \emph{Swift}-XRT data started to be collected after optical maximum light.

Finally, XMM-Newton observed ASASSN-15lh on November 18, 2015 ($\delta t=134.2$ days rest-frame since maximum light), six days after our first CXO epoch, which yielded a non-detection. From the XMM-Newton observations \citet{leloudas16} infer a $95\%$ confidence level flux limit $F_x<2\times 10^{-16}\,\rm{erg\,s^{-1}cm^{-2}}$ (0.3-1 keV). We do not confirm the results from \citet{leloudas16}. Adopting their inferred count-rate limit of 11 source counts in 9 ks of exposure time with EPIC-MOS2, their assumed blackbody spectrum with $T=18$ eV, and following the flux calibration procedure outlined in \citet{leloudas16}, we infer a flux limit which is $\sim100$ times larger. 

We re-analyzed the XMM data using standard routines in the Scientific Analysis System (SAS version 15.0.0) and the relative calibration files. We employ a source region of 32$\arcsec$ radius and extract the background from a source-free region on the same chip. No X-ray source is detected at the location of ASASSN-15lh. Our best constraints are derived from observations obtained with EPIC-MOS2, with total exposure time of 9 ks (after removal of time windows contaminated by proton flaring) and a 3$\,\sigma$ count-rate upper limit of $0.002\,\rm{c\,s^{-1}}$ (0.3-10 keV).  For the best fitting spectral models derived from CXO detections, we infer the following unabsorbed 0.3-10 keV flux limits: $F_x<1.4\times 10^{-14}\,\rm{erg\,s^{-1}cm^{-2}}$ and $F_x<1.5\times 10^{-14}\,\rm{erg\,s^{-1}cm^{-2}}$ for the blackbody and the power-law spectrum, respectively. XMM observations do not reach the necessary depth to probe the emission from the X-ray source that we detect with CXO and the stacking of \emph{Swift}-XRT observations. A summary of the results from the X-ray observations of ASASSN-15lh can be found in Table \ref{Tab:Xray}.

%%%%%%%%%%%%%%%%%%%%%%%%%%%%%%%%%%%%
\subsection{UV analysis}
\label{SubSec:UV}

\begin{figure}
\vskip -0.0 true cm
\centering
\includegraphics[scale=0.55]{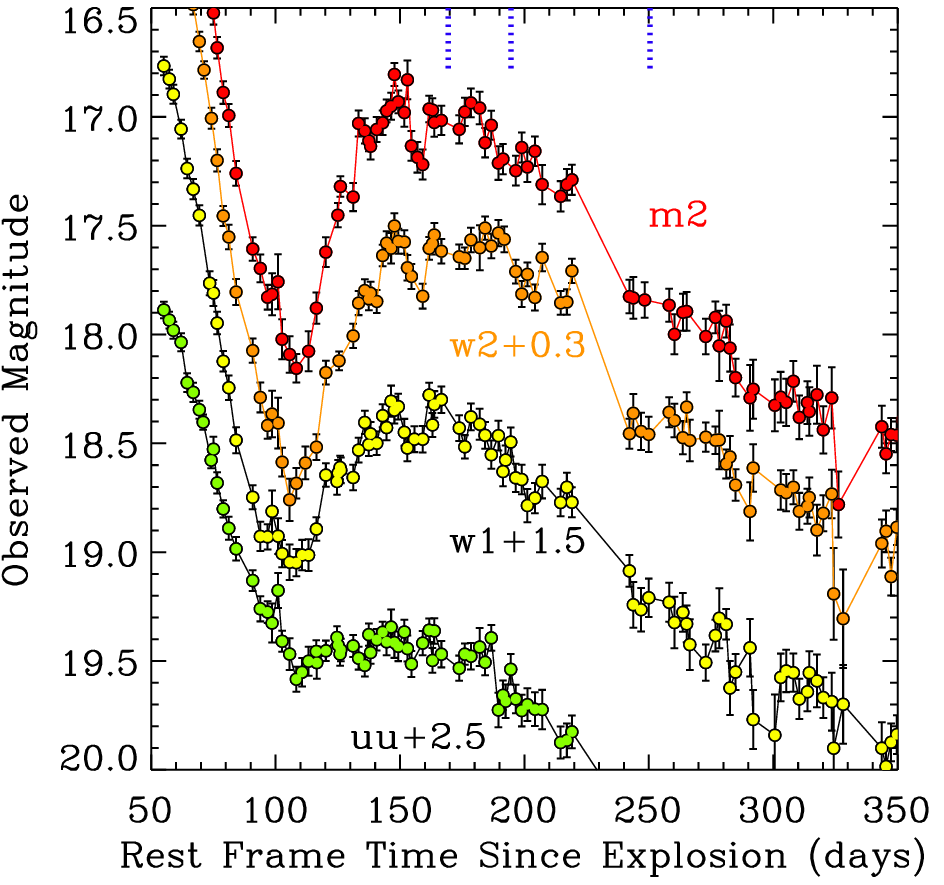}
\caption{Swift-UVOT observations of ASASSN-15lh reveal the presence of pronounced time variability on short time scales $\Delta t \lesssim 5$ days across the \emph{Swift}-UVOT bands at the time of the re-brightening. Vertical dotted lines mark the times of the CXO observations. }
\label{Fig:UVvar}
\end{figure}

We re-analyzed all the \emph{Swift}-UVOT observations obtained from June 24, 2015 until July 22, 2016 following the prescriptions by \cite{Brown09} and adopting the updated calibration files and revised zero points by \cite{Breeveld11}. Each individual frame has been visually inspected and quality flagged. Observations with insufficient exposure time have been merged  to obtain higher signal-to-noise ratio (S/N) images from which we extracted the final photometry reported in Table \ref{Tab:UVOT}. We corrected for Galactic extinction in the direction of the transient  ($E(B-V)=0.03$ mag, \citealt{Schlafly11}) and subtracted the host galaxy flux component as constrained by \cite{Dong16}. We performed a self-consistent flux calibration, and applied a dynamical count-to flux conversion that accounts for the spectral evolution of ASASSN-15lh, following the procedure outlined in \cite{Margutti14}. Finally, we computed a bolometric light-curve of ASASSN-15lh by integrating the best-fitting blackbody spectra. 

A partial collection of the \emph{Swift}-UVOT photometry of ASASSN-15lh has already been presented by \cite{Dong16}, \cite{Godoy-Rivera16}, \cite{Brown16} and \cite{leloudas16}. Here we update the observations and focus on the presence of significant temporal variability that appears at the time of the re-brightening. Figure \ref{Fig:UVvar} shows the presence of pronounced temporal variability across the UVOT bands, and more pronounced at UV wavelengths as first noticed by \cite{Brown16}. The short variability time scale $\Delta t \lesssim 5$ days at $\sim150$ days since first light argues against the interpretation of the SN shock interaction with the surroundings as the main source of energy powering the re-brightening \citep{Chatzopoulos16}. For a typical SN shock velocity $v_{sh}\sim0.2$ c (e.g. \citealt{Margutti14b}, their Fig. 2) we do not expect significant temporal variability on $\Delta t<30$ days at $t\sim150$ days, contrary to what we observe in ASASSN-15lh. This observation motivates us to consider alternative explanations of the UV re-brightening (Sec. \ref{SubSec:Rep}).

%%%%%%%%%%%%%%%%%%
\subsection{Late-Time Optical Spectroscopy}
\label{SubSec:Spec}

\begin{figure*}
\vskip -0.0 true cm
\centering
\includegraphics[width=7.2in]{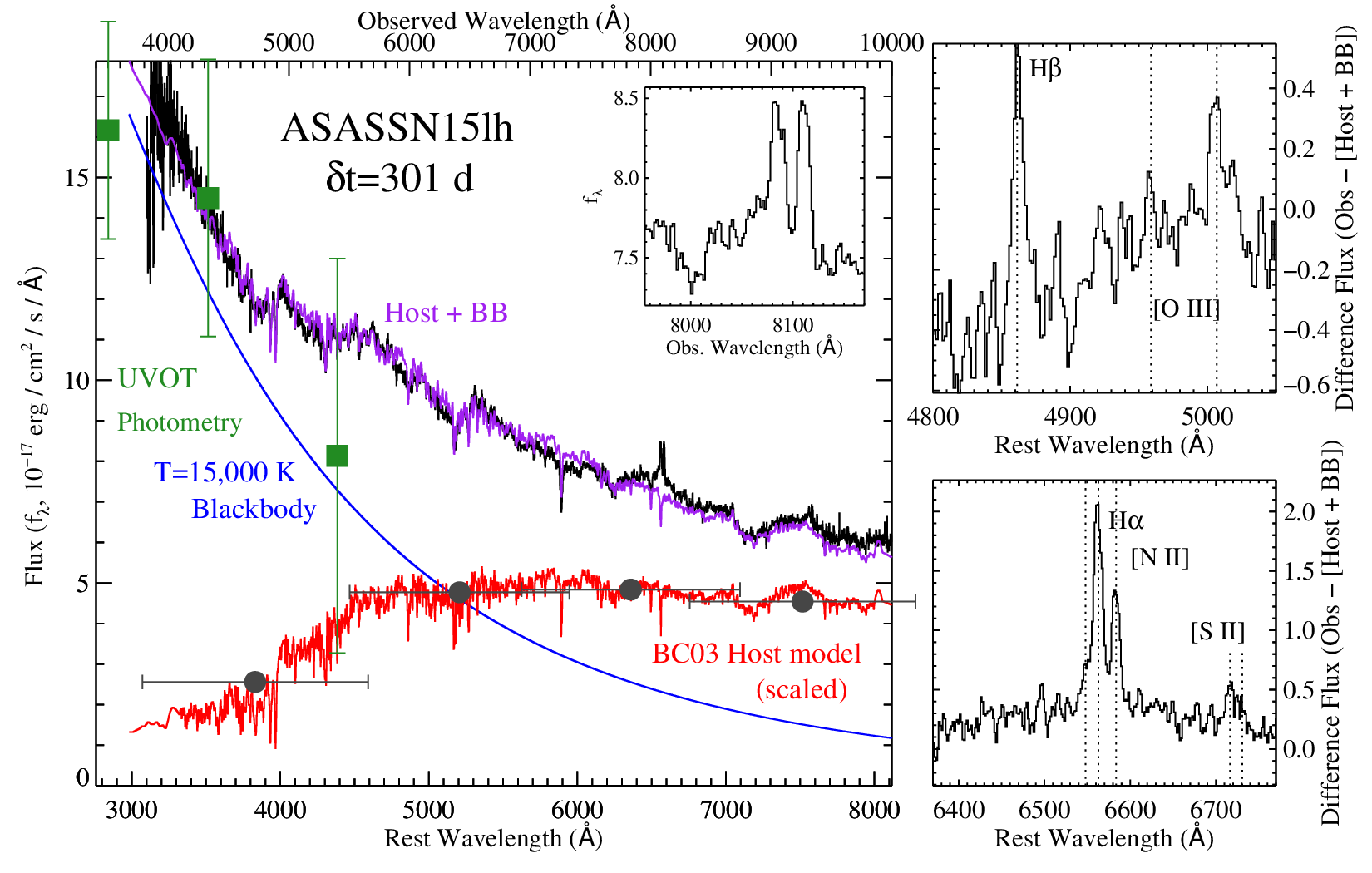}
\caption{\emph{Left Panel}: A decomposition of the observed late-time spectrum (black line) obtained on 2016 June 10 ($\delta t$= 301 days rest-frame since maximum light) into a sum (magenta line) of the FAST host-galaxy model that best fits the pre-transient photometry (red line) and a featureless blackbody with temperature inferred from our fits of ASASSN-15lh (blue line), demonstrates that the strongest spectral features can be reasonably attributed to the underlying host galaxy stellar population. The green squares mark the {\it Swift} UVOT photometry after subtraction of the host model and interpolated to the date of observation. The gray circles are the scaled values for the broadband photometry of the host \citep{Dong16}.  We emphasize the lack of unambiguous evidence for broad spectral features associated with the transient. \emph{Inset}: The observed spectrum in the vicinity of the H$\alpha$/[\ion{N}{2}] complex is shown enlarged.  Two clear emission lines are present prior to any correction for the stellar continuum.
\emph{Right Panels}: Zoom-ins to spectral regions of interest in the difference spectrum show the presence of \emph{narrow} emission lines at the expected wavelengths (dotted lines) associated with H$\alpha$, H$\beta$, [\ion{N}{2}] $\lambda\lambda$6548, 6583, and possibly [\ion{S}{2}] and [\ion{O}{3}]~$\lambda$5007 at a common redshift of $z$=0.2320. }
\label{Fig:spec1}
\end{figure*}

We acquired deep multi-epoch optical spectroscopy of ASASSN-15lh, spanning the time range $\delta t=$ 35--350 rest-frame days after maximum light and sampling key points in the late evolution of the transient. A more detailed analysis will be presented in future work (Chornock et al., in prep.). Here we concentrate on an analysis of our highest S/N late-time spectrum, which was acquired well after the second re-brightening and when the underlying emission from the host galaxy stellar population is better revealed. 

We observed ASASSN-15lh on 2016 June 10 ($\delta t=301$ days rest-frame since maximum light) using the Low Dispersion Survey Spectrograph (LDSS3C; \citealt{ldssref}) on the 6.5~m Magellan Clay telescope.  We obtained three 1800~s exposures using the VPH-All grism and a 1\arcsec-wide slit near the center of the field of view oriented at a position angle of 128\fdg3, which was close to the parallactic angle  \citep{fil82}. This setup covered the range 3800--10500~\AA\ with a resolution of 8.1~\AA. Standard IRAF\footnote{IRAF is
  distributed by the National Optical Astronomy Observatory,
    which is operated by the Association of Universities for Research
    in Astronomy, Inc., under cooperative agreement with the National
    Science Foundation.} tasks were used to perform two-dimensional image processing.  We used custom IDL scripts to perform flux calibration and correction for telluric absorption using observations of EG131 obtained immediately prior to the object. We took particular care to mitigate the effects of second-order light contamination by combining observations of the standard star taken both with and without an order-blocking filter.  However, small residual contamination at long wavelengths ($\lambda$>8000~\AA) is possible.

The resulting spectrum is shown in black in Figure~\ref{Fig:spec1}.  Numerous stellar absorption features from star light in the host galaxy are visible, as well as two emission peaks near H$\alpha$ (observed wavelengths $\sim$8100~\AA).  Several authors have fit the available pre-outburst host galaxy photometry \citep{melchior15,Dong16,leloudas16} and have found consistent results.  However, the presence of spectral features from the host stellar population has the potential to improve the constraints on the stellar population synthesis, so we used an iterative procedure to incorporate this information while avoiding the flux from the transient.

First, we estimated a best-fit blackbody temperature of $T_{\mathrm{BB}}\approx15,000$~K at the time of observations from the analysis of the UVOT photometry described above.  We then subtracted a scaled blackbody spectrum from the observed spectrum under the constraint that the blackbody-subtracted spectrum had to match the observed colors of the host galaxy to obtain an initial estimate of the host-only spectrum.
We then used the FAST code \citep{kriek09} to fit the host-only spectrum combined with, and normalized by, the broadband $grizyJK$ host photometry \citep{melchior15,Dong16}.  For simplicity, we fixed the metallicity to solar and assumed a \citet{chabrier03} initial mass function and zero internal extinction.  We obtained a satisfactory fit using the \citet{bc03} stellar models and an exponentially-declining star formation law.  The best-fit model has a total stellar mass of 1.2$\times$10$^{11}$~$M_{\sun}$, a current stellar age of 10~Gyr, and an $e$-folding timescale of 2~Gyr, resulting in a current star-formation rate of $\sim$0.8~$M_{\sun}\,\rm{yr^{-1}}$.  These numbers are in broad agreement with those reported previously (e.g, \citealt{Dong16,leloudas16}).
Other choices for the stellar population model produced qualitatively similar results, although usually with smaller current star-formation rates.  Our best fit for the host is plotted in red in Figure~\ref{Fig:spec1}. 

We then fitted our observed spectrum as a linear combination of the host galaxy model and a blackbody to find appropriate flux scaling factors. The scaled blackbody is plotted as blue in Figure~\ref{Fig:spec1} and good agreement can be seen with the host-subtracted UVOT $ubv$ photometry (green squares) interpolated to the date of observation. Both the fitted host spectrum and the overplotted host photometry (gray circles) have been scaled by a factor of 0.40 from the values for the whole host, which presumably results from the smaller size of our spectroscopic aperture relative to the host as a whole.  The \citet{bc03} models clearly have narrower features than those visible in our spectrum, so the host template was smoothed with a 10~\AA\ boxcar function to mimic the combined effects of our spectral resolution and the internal velocity dispersion of the host galaxy.  Our results are not very sensitive to the width of this smoothing kernel.  The sum of the scaled blackbody and the smoothed galaxy template is plotted in magenta in Figure~\ref{Fig:spec1} and is a good match to the observed spectrum in black.

%\textcolor{red}{Brief description of data acquisition and reduction,  modeling of the host.}
\citet{leloudas16} noted two emission peaks near 4000 and 5200~\AA\ in their late-time spectra of ASASSN-15lh.  However,
accurate modeling of the host galaxy stellar component from our late-time spectrum demonstrates that the most prominent broad spectral features detected in the observed (host plus transient) late-time spectra have to be attributed to the underlying continuum from the host galaxy star light (Figure \ref{Fig:spec1}). We do not find unambiguous evidence for broad spectral features associated with the transient at this epoch. Small, broad, low-amplitude discrepancies between the observed spectrum and combined fit (black and magenta lines, respectively) are present, but it is not yet clear if they represent true spectral features of the transient or limitations in the stellar population synthesis modeling. 
More observations of the host will be required after the optical transient fades further to more accurately constrain the presence of possible low amplitude broad spectral features in the transient spectrum at late times.

Without any correction for the host galaxy, the spectrum has the two obvious narrow emission features near 8085 and 8111~\AA\ (in air) noted above, which can be clearly associated with H$\alpha$ and [\ion{N}{2}]~$\lambda$6583 at $z$=0.2320 (lower-right panel of Figure~\ref{Fig:spec1})\footnote{This redshift is consistent with that measured from the stellar absorption features.  Note that this value is slightly offset from the redshift $z$=0.2326 measured from narrow UV absorption lines \citep{Dong16,Brown16,leloudas16}. We do not discuss further the implications of this possible velocity offset for the UV absorbers in this work.}.  [\ion{N}{2}]~$\lambda$6548 is blended in the blue wing of H$\alpha$.  H$\beta$ is only visible in emission after subtraction of the host model.
%Instead, as the transient faded, narrow emission lines associated with H$\alpha$, H$\beta$, and [\ion{N}{2}] became more prominent.  
Weaker features also appear to be present in the difference spectrum near the [\ion{S}{2}] doublet and [\ion{O}{3}]~$\lambda$5007. We searched for [\ion{O}{2}] $\lambda$3727 emission and none is visible, but the S/N of the spectrum is not as high at those wavelengths.  Inspection of our spectral sequence reveals that the H$\alpha$/[\ion{N}{2}] lines are present in several of our higher S/N spectra throughout the evolution of the transient, consistent with a constant low-level contribution that is strongly diluted by light from the transient at earlier times.

%\bluepen{line ratios, narrow lines,...}
The peaks of H$\alpha$ and [\ion{N}{2}]~$\lambda$6583 are of comparable height \emph{prior} to subtraction of the host model.  Strong [\ion{N}{2}]/H$\alpha$ is a possible sign of ionization by an AGN-like continuum.  However, after correction for the underlying Balmer absorption in our best-fit host model, the ratio decreases to $\sim$0.5.
This line ratio, combined with weak [\ion{O}{3}]/H$\beta$ and [\ion{S}{2}]/H$\alpha$, is consistent with the nebular emission being powered by star formation instead of AGN activity \citep{kewley06}. We caution that these ratios are sensitive to systematic errors in the modeling of the underlying stellar absorption, and in particular the strength of the stellar Balmer absorption.
If all of the inferred H$\alpha$ emission (flux $\sim$2.7$\times$10$^{-16}$ erg~cm$^{-2}$~s$^{-1}$) is powered by star formation, the inferred rate is $\sim$0.4~$M_{\sun}$~yr$^{-1}$ \citep{Kennicutt98}, in rough agreement with that estimated from the host galaxy stellar population fit.

\citet{leloudas16} reported H$\alpha$ emission from ASASSN-15lh with a full width at half-maximum (FWHM) of 2500~km~s$^{-1}$, but in our data, it is clear that the reported emission feature is just the narrow nebular H$\alpha$ and [\ion{N}{2}] from the host blended together at low S/N or low resolution in their data. Note that in their highest S/N spectra (inset of their Figure 1), the putative H$\alpha$ from the transient is flat-topped or double peaked, consistent with the two strong nebular emission lines of roughly equal height (inset of our Figure~\ref{Fig:spec1}) being blended together.  We also note that \citet{leloudas16} do not attempt to correct for the contribution from the underlying stellar continuum. Therefore, we do not confirm their claim of H$\alpha$ emission from the transient itself and the reported velocity FWHM likely reflects the spacing of the two [\ion{N}{2}] lines, which are each offset by $\sim$1000~km~s$^{-1}$ from the central H$\alpha$ emission.  \citet{Godoy-Rivera16} also report a "bump" near H$\alpha$ at late times, but they do not report a FWHM, so it is not clear if they are are also possibly referring to a noisy detection of the narrow nebular lines.

%%%%%%%%%%%%%%%%%%
\subsection{Re-analysis of Early-Time Optical Spectra}
\label{SubSec:Spec2}

\begin{figure}
\centering
\includegraphics[width=\linewidth]{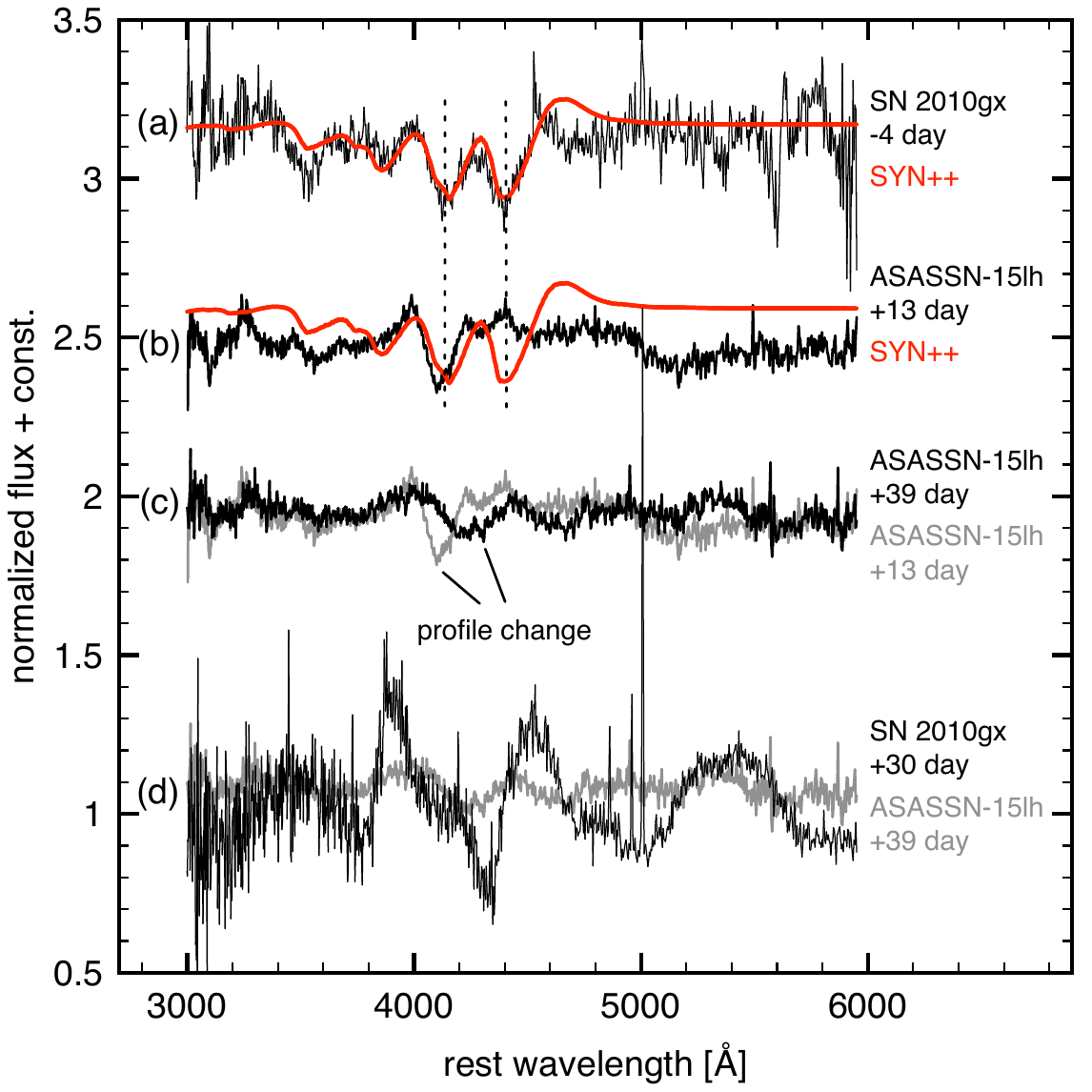}

\caption{Placing early-phase optical spectra of ASASSN-15lh in context with SLSNe. (a) The \ion{O}{2} ion is a signature of SLSNe, and here we show how using the simple assumptions of SYN++ and a photospheric velocity of 19,000 km~s$^{-1}$, the -4 day spectrum of the SLSN 2010gx can be reproduced. (b) By contrast, we cannot reproduce +13 day spectrum of ASASSN-15lh. It clearly misses an accompanying feature around 4400~\AA. (c) Evolution in the spectra are observed. Most conspicuous is the 4100~\AA\ feature that drifts to longer wavelengths. (d) The evolution toward increasingly inconspicuous spectral features is unlike SLSNe that exhibit increasingly stronger spectral features. Here we show the +30 day spectrum of SN\,2010gx, which is unlike the nearly featureless +39 day spectrum of ASASSN-15lh. Data have been retrieved from WISEREP \citep{YG12}, normalized according to the procedure outlined in \citet{Jeffery07} to aid in visual comparison, and were originally published in \citet{Pastorello10} and \citet{Dong16}.}

\label{fig:spec-compare}
\end{figure}

In addition, we re-evaluated the early optical spectra of ASASSN-15lh and were unable to confirm the likeness to SLSNe reported by \citet{Dong16}. The  \ion{O}{2} ion, which is commonly observed in SLSNe \citep{Quimby11}, has a number of distinctive absorption features not observed in ASASSN-15lh (Fig. ~\ref{fig:spec-compare}). The strongest two features centered near 4100 and 4400~\AA\ are always observed to be of comparable strength and no reasonable values of temperature or density can change this ratio. ASASSN-15lh only shows the 4100~\AA\ feature (Fig. \ref{fig:spec-compare}, see also \citealt{leloudas16}).  Without the accompanying 4400~\AA\ feature, it is hard to reconcile the proposed association with \ion{O}{2}, and thus the spectroscopic connection to SLSNe is not robust.

The spectral features of ASASSN-15lh trend redward over time toward declining velocities. This is similar to the spectral evolution of supernovae where it is attributed to an expanding and cooling photosphere. However, unlike supernovae, the features of ASASSN-15lh do not show traditional P-Cyg profiles and become increasingly inconspicuous. For example, the +30 day spectrum of SN\,2010gx exhibits pronounced features, whereas the +39 day ASASSN-15lh spectrum is nearly featureless (Fig. \ref{fig:spec-compare}). We explored a variety of  possible ions using the highly parameterized spectrum synthesis tool, \texttt{SYN++} \citep{Thomas11}, to determine whether blending of features could reproduce the spectral features and evolution of ASASSN-15lh, but were unsuccessful.

To our knowledge the only previous examples of spectral features becoming increasingly inconspicuous in the early phases of a supernova involve interaction with dense CSM. SN-CSM interaction can rescale or ``mute'' the line profile relative to the continuum \citep{Branch00}. Most confirmed instances of SN-CSM interaction involve H-rich material that can be readily identified by the presence of H Balmer lines that may be narrow ($<$ 100 km\,s$^{-1}$) to broad ($\sim 10^3$ km\,s$^{-1}$) in width, depending on their origin of formation. The hydrogen-poor SLSN iPTF13ehe exhibited H$\alpha$ Balmer emission with broad and narrow components +251 days post maximum \citep{Yan15}. However, no such lines are observed in ASASN-15lh. Interaction with H-poor CSM ejected by rapidly rotating pulsational pair instability supernovae is possible \citep{Chatzopoulos12,Chatzopoulos16}, but the spectroscopic consequences of such interaction are poorly understood \citep{Chatzopoulos13}, and the timescales of variability observed in the UV strongly favor against this scenario (section \ref{SubSec:UV}).

A luminous central source over-ionizing expanding ejecta is a speculative, though attractive, scenario that may explain the spectroscopic evolution of ASASSN-15lh toward a featureless continuum. As there is no precedent for this scenario, the specific spectral signatures are unclear. Certainly, the ionizing photons must be extremely energetic for no strong optical or UV lines to be observable. An analogous phenomenon may be variable UV absorption commonly seen in Seyfert galaxies \citep{Maran96,Crenshaw00}. In some cases variability in the form of absorption components appearing and disappearing, or decreasing outflow velocities \citep{Gabel03}, can result from changes in the ionizing flux \citep{Kraemer02}. ASASSN-15lh may be an extreme version of these processes.

%%%%%%%%%%%%%%%%%%%%%%%%%%%%%%%%%%%%%%%%%%%
\section{Interpretation}
\label{Sec:Int}

\begin{figure}
\vskip -0.0 true cm
\centering
\includegraphics[width=.5\textwidth]{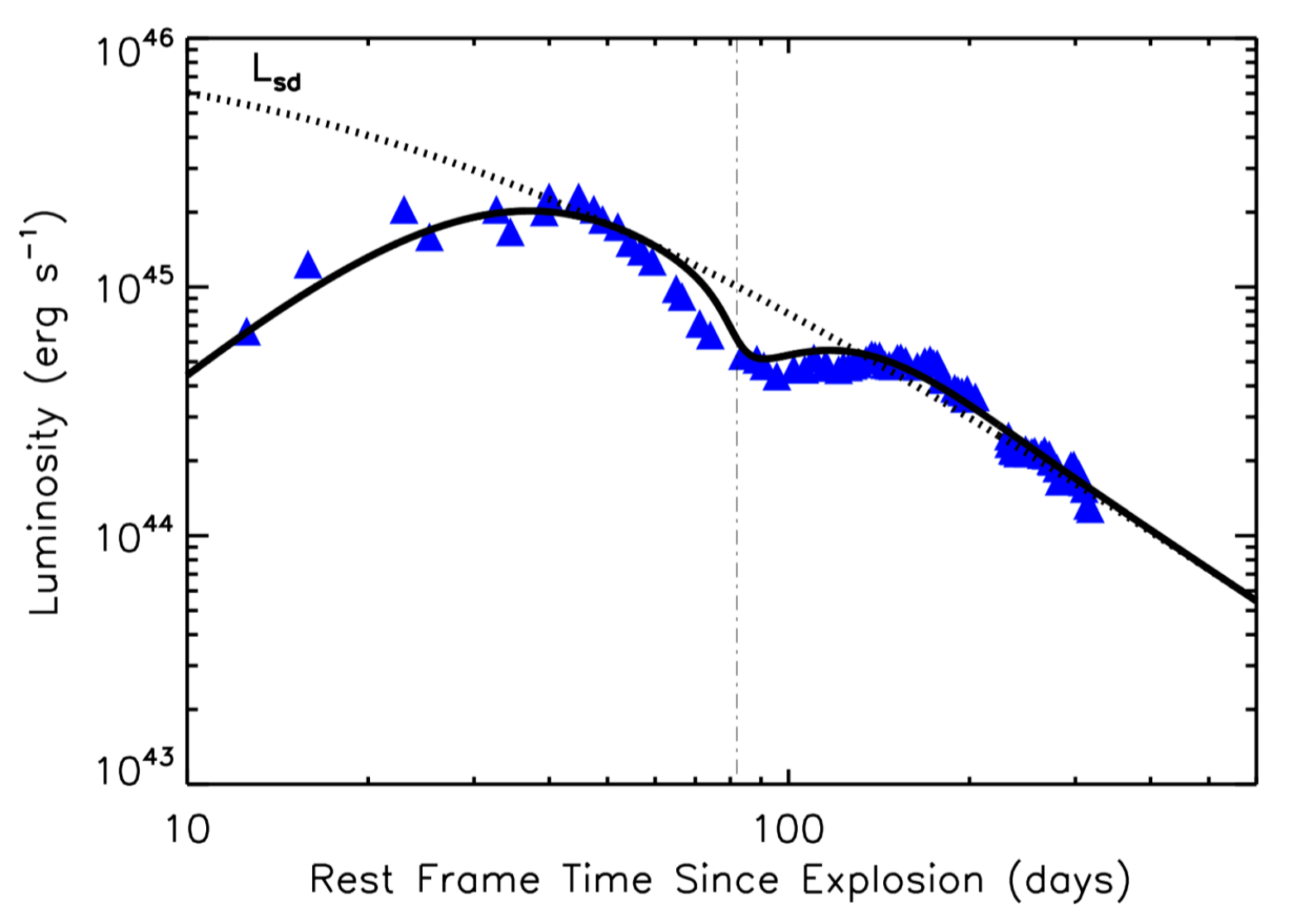}
\caption{Bolometric light curve model (solid line) due to reprocessed luminosity from a central source (dashed line). For this plot we use the spin down power from a magnetar engine as central source as an example. We artificially and abruptly change the ejecta opacity at the time shown by the vertical dashed line to approximate the effect of an ionization break-out.  The object's time of first light is assumed to occur 30 days before the light curve peak.  The bolometric light curve of ASASSN-15lh that we derived in Sec. \ref{SubSec:UV} is shown for comparison with blue triangles. }
\label{Fig:Opacity}
\end{figure}

%----------------------------------------------------------------------------------------------------------------
\subsection{The ``Reprocessing Picture"}
\label{SubSec:Rep}
Although the mechanisms behind SLSNe powered by a stellar-mass compact object$-$such as a magnetar$-$ and the tidal disruption and accretion of a star by a SMBH do differ significantly, the basic physical process driving the light curves of these events may be similar.
A central source of UV/X-ray radiation (an accreting SMBH or the pulsar wind nebula of a rapidly spinning neutron star, NS) is absorbed by a dense column of gas, and downgraded into optical radiation, where the lower opacity allows the radiation to more readily escape.
Such a ``reprocessing" picture has been applied to explain both TDEs (\citealt{Loeb97,Guillochon13,Metzger16}) and SLSNe (\citealt{Kasen10,Woosley10, Metzger14}).

Consider the characteristic timescale of the central engine in a magnetar-powered SLSN and TDE scenarios.  In a magnetar-powered SLSN, the central engine lifetime is the magnetic dipole spin-down timescale of the magnetar:
\begin{equation}
t_{\rm sd} \simeq 17\,{\rm d}(M/1.4M_{\odot})^{3/2}B_{13}^{-2}P_{\rm ms}^{2},
\label{eq:tsd}
\end{equation}
where $M$, $P = P_{\rm ms}$ ms, and $B_{\rm d} = 10^{13}B_{13}\,{\rm G}$ are, respectively, the mass, initial spin period, and dipole surface magnetic field strength of the magnetar (e.g.~\citealt{Spitkovsky06}).  The maximum energy of the engine is limited to the rotational energy of the NS,
\begin{equation}
E_{\rm rot} = I\Omega^{2}/2 \simeq 2.5\times 10^{52}{\rm erg}\,\,(M/1.4M_{\odot})^{3/2}P_{\rm ms}^{-2}\,,
\label{eq:Erot}
\end{equation}
which can vary from $\sim 3\times 10^{52}-10^{53}$ erg for the minimum value of the spin-period set by the mass shedding limit, depending on the mass of the NS (\citealt{Metzger15}).  
In order to simultaneously explain the large radiated energy and duration of ASASSN-15lh with a magnetar, we require a maximally spinning neutron star ($P_{\rm ms} \lesssim 1$) and a relatively weak magnetic field $B_{\rm d} \sim 10^{12}-10^{13}$ G (e.g., \citealt{Metzger15}, see also \citealt{Dong16,Chatzopoulos16, Bersten16, Sukhbold16}).   

In the TDE scenario, the engine lifetime is uncertain, but is commonly attributed to the fall-back time of the most tightly bound stellar debris following the disruption (e.g., \citealt{Guillochon13,Stone13}),
\begin{equation}
t_{\rm fb} \approx 410\,{\rm d}\left(\frac{M_{\bullet}}{10^{8}M_{\odot}}\right)^{1/2}\left(\frac{m_{\star}}{M_{\odot}}\right)^{1/5},
\label{eq:tfb}
\end{equation}
where $M_{\bullet}$ and $m_{\star}$ are the mass of the SMBH and the star, respectively, and we have assumed a stellar mass-radius relationship $R_{\star} \simeq R_{\odot}(m_{\star}/M_{\odot})^{4/5}$ appropriate to lower main-sequence stars.  The maximum radiated energy is that liberated by the accretion of the half of the stellar mass which remains bound to the SMBH,
\begin{equation}
E_{\rm acc} = \eta (m_{\star}/2)c^{2} \simeq 9\times 10^{52}\,{\rm erg}\left(\frac{\eta}{0.1}\right)\left(\frac{m_{\star}}{M_{\odot}}\right),
\end{equation}
where the radiative efficiency for geometrically thin accretion varies from $\eta \approx 0.04-0.42$, depending on the spin of the SMBH and its orientation relative to the angular momentum of the accreting gas. 

In the TDE scenario, the energetics of ASASSN-15lh are reasonably accommodated by the accretion of a solar-mass star.  However, the high mass SMBH $M_{\bullet} \sim 6\times 10^{8}M_{\odot}$ inferred from the host of ASASSN-15lh, would appear to predict a long duration of the transient $t_{\rm fb} \gtrsim$ 2 year, inconsistent with the much shorter observed decay time of the first peak of a few weeks. This inconsistency could be solved by considering that a main-sequence star can only be disrupted by such a massive black hole if the SMBH is \emph{spinning} in a prograde direction with respect to the orbit of the disrupted star (e.g. \citealt{Kesden12}).  Precession of the star during the phase of tidal compression due to the BH spin may substantially enhance the spread in the energy distribution of the stellar debris as compared to the Newtonian case, by partially aligning the direction of the hydrodynamic bounce with the velocity vector of the star (\citealt{Metzger16,leloudas16}).  More tightly bound debris has a shorter orbital period, which could significantly speed-up the flare evolution timescale as compared to the Newtonian gravity estimate in Equation \ref{eq:tfb}.  Though promising, general relativistic numerical simulations are needed to confirm this possibility.

In addition to possibly speeding up the flare evolution, the high BH spin required to explain ASASSN-15lh as a TDE would (i) naturally result in a large value of the accretion efficiency $\eta$, accounting for its high luminosity; (ii) possibly aid in the process of debris circularization by inducing precession of the stellar debris streams (e.g.~\citealt{Dai15}, \citealt{Hayasaki+16}, \citealt{Bonnerot+16}).  Precession of the streams out of the orbital plane due to misaligned BH spin could also help make the geometry of the reprocessing material relatively spherical (e.g., \citealt{Guillochon15}), consistent with the low measured optical polarization of ASASSN-15lh reported by \cite{Brown16}. 

%----------------------------------------------------------------------------------------------------------------
\subsection{X-rays Escape and the Ionization Break-Out}
In both the magnetar SLSN and TDE scenarios, UV/X-ray radiation from the central source may ionize its way through the ejecta at late times.  This process can result in the direct escape of UV/soft X-ray radiation while having an indirect influence on the observed optical light curve by changing the ejecta opacity (Sec. $\ref{SubSec:LC}$).  

Approximating the ejecta as a homogeneously expanding sphere of mass $M_{\rm ej}$, velocity $v_{\rm ej} = 10^{9}v_{9}$ cm s$^{-1}$, and radius $R_{\rm ej} = v_{\rm ej}t$,  the neutral column density is
\begin{equation}
\Sigma (t)\simeq \frac{M_{\rm ej}}{(4\pi/3)R_{\rm ej}^{2}m_p} \approx 1.7\times 10^{24}\,{\rm cm^{-2}} f_n\left(\frac{M_{\rm ej}}{M_{\odot}}\right)v_9^{-2}\left(\frac{t}{150{\rm d}}\right)^{-2},
\end{equation}
where $f_n$ is the neutral fraction.  This is much higher than the inferred X-ray absorption column of $NH_{int}< 3\times 10^{22}$ cm$^{-2}$ towards ASASSN-15lh, requiring a very low neutral fraction if the X-ray source is related to the optical transient. This is consistent with the very low N(HI) inferred by \cite{leloudas16} from Ly-$\rm{\alpha}$.

The ejecta from TDEs and SLSNe are expected to have markedly different chemical composition. In a TDE the ejecta has nearly solar composition (e.g.~\citealt{Kochanek16a}) and the escape of soft X-rays is inhibited primarily by the bound-free opacity of neutral helium (\citealt{Metzger16,Roth15}).  By contrast, in a H-poor SLSN, X-rays are blocked more severely by neutral oxygen and carbon (\citealt{Metzger14}).  

A central engine with an UV/X-ray luminosity $L$ releases an energy $L \times t$ in ionizing radiation on a timescale $t$.  If the ejecta contains a mass fraction $X_Z$ of elements with atomic number $Z = 8Z_8$, the radiation ionizes its way through the ejecta on a timescale 
\begin{eqnarray}
t_{\rm ion} \approx  \left\{
\begin{array}{lr}
120\,{\rm d}\,M_3^{3/4} v_9^{-5/4}T_{5}^{-0.2}\left(\frac{X_{A}}{0.1}\right)^{1/4}\left(\frac{L t}{10^{52}{\rm ergs}}\right)^{-1/4}Z_8^{3/4}
, 
(\eta_{\rm thr} \ll 1) \\
110\,{\rm d}\,M_3 v_9^{-3/2}T_{5}^{-0.4}\left(\frac{X_{A}}{0.1}\right)^{1/2}\left(\frac{L t}{10^{52}{\rm ergs}}\right)^{-1/2}Z_8^{3/2},       
(\eta_{\rm thr} \gg 1), \\
\end{array}
\right.
\label{eq:tbo}
\end{eqnarray}
where $M_3 \equiv M_{\rm ej}/(3M_{\odot})$ and $T_{5} = T/10^{5}$ K is the temperature of electrons in the recombination layer and
\begin{eqnarray}
\eta_{\rm thr} \approx 0.7\left(\frac{L t}{10^{52}{\rm erg}}\right)^{-1}M_{3}v_9^{-1}\left(\frac{X_{A}}{0.1}\right)T_{5}^{-0.8}Z_8^{3}
\label{eq:etaA}
\end{eqnarray}
is the ratio of absorptive and scattering opacity in the ejecta \citep{Metzger14}. 

For typical parameters and an engine similar to ASASSN-15lh with $L t \sim 10^{52}$ ergs, we have $t_{\rm ion} \sim $ month in the case of a He-rich composition ($Z = 2$) of a TDE-like scenario.  By contrast, for a CO-rich composition of an exploding massive star ($Z = 8$), we have $t_{\rm ion} \sim $ months, making break-out harder to achieve.  In the latter case, X-ray break-out is even less likely considering that the K-shell valence electrons of oxygen have a binding energy of $\sim 1$ keV, while the {\it measured} keV X-ray luminosity of ASASSN-15lh $\leq 10^{42}$ erg s$^{-1}$ is much less than the optical/UV luminosity (in other words, the true value of $L t$ to use in Equation \ref{eq:tbo} should be much lower than $10^{52}$ erg).

We conclude that an ionization break-out could allow the escape of X-rays in the TDE scenario, but is probably not sufficient to do so in the case of a H-poor supernova given the observed soft X-ray spectrum. % \textcolor{red}{Comment on tau-free free, radio is suppressed consistent with obs}

%\textcolor{red}{For the massive star interpretation we can use the X-rays to put a limit to the close environment, hence Mdot for massive stars. Vague about the origin of the X-rays in this case.}
%----------------------------------------------------------------------------------------------------------------
\subsection{The double-humped light-curve of ASASSN-15lh}
\label{SubSec:LC}

The ionization of the ejecta reduces the bound-free opacity, allowing the escape to the observer of UV and X-ray radiation with energies \emph{above} the ionization threshold. This process is unlikely to explain the observed UV re-brightening by itself, as even the highest frequency UV bands of \emph{Swift}-UVOT are below the first ionization energies of the most abundant elements (H, He, C, O). However, an ionization break-out may have an indirect effect on the light-curve via the continuum opacity.
 
At early times the  ejecta is largely neutral and the opacity at optical frequencies is dominated by electron scattering, while the opacity at UV frequencies is dominated by line transitions of metals. However, once the ejecta becomes ionized by the central engine, the electron scattering opacity will increase, while the UV opacity will decrease as the ionized atoms have fewer bound-bound transitions.  Therefore, following ionization break-out we expect a shift of the peak of the spectral energy distribution from optical to UV frequencies. The appeal of  this model is that a \emph{single} central-engine timescale would naturally reproduce the double-peaked temporal structure of ASASSN-15lh, which has no analogue in previously observed TDE or SLSN light curves. As a comparison, the TDE model invoked by \cite{leloudas16} combines two luminosity mechanisms, which result into two different timescales. 

While accurate modeling, beyond the scope of this paper, is necessary to understand if this effect alone can quantitatively explain the observations of ASASSN-15lh, here we consider a toy model to illustrate the basic principles. For illustrative purposes we use the spin down luminosity of a magnetar as the central source of ionizing photons. In particular, 
we consider a magnetar light-curve with parameters $P = 1$ ms and $B_{\rm d} = 3\times 10^{12}$ G, similar to that described in \cite{Metzger15}, and a total ejecta mass of $M_{\rm ej} = 10M_{\odot}$.  However, we artificially change the grey opacity from $\kappa_i = 0.02$ cm$^{2}$ g$^{-1}$ to $\kappa_f = 0.2$ cm$^{2}$ g$^{-1}$ at a time corresponding to ionization break-out of about 50 days.  As shown in Fig. \ref{Fig:Opacity}, this produces a minimum/flattening in the bolometric light curve, similar to that observed in ASASSN-15lh.  Although we have applied the model to a magnetar for concreteness (and since the process of debris circularization in TDEs remains uncertain), a similar result applies to the TDE case if the central UV/X-ray accretion power smoothly rises on a timescale of a few weeks and then decays $\propto t^{-5/3}$ at later times.  We also caution that a simple change in the grey opacity is unlikely to accurately predict the effect of wavelength-dependent opacity change created by an ionization break-out.  
%----------------------------------------------------------------------------------------------------------------
%\subsection{Considerations on the intrinsic nature of the ionizing source that powers ASASSN-15lh}
%So we can basically explain the phenomenology with a light-bulb and some material around. If we force 15lh into the context of known things, then two possibilities come to mind. Explore the implications, like weird star formation going on close to the nucleus leading to stellar progenitors with weird properties, and maybe this is connected to the low level of activity of the galaxy?  How we produce a magnetar with those properties?
%%%%%%%%%%%%%%%%%%%%%%%%%%%%%%%%%%%%%%%%%%%
\section{Summary and Conclusions}
 \label{Sec:conclusions}
We presented evidence for luminous, soft and persistent X-ray emission at the location of ASASSN-15lh, and discussed its origin in the context of multi-wavelength observations of the transient, which include constraints on its radio emission and early and late-time optical spectroscopy. Our re-analysis of early-time spectra does not confirm the robust association of ASASSN-15lh with SLSNe claimed by previous studies, and invites us to be open-minded about the nature of ASASSN-15lh. Late-time spectra reveal the emergence of \emph{narrow} emission features from the host galaxy, while we associate the most prominent  \emph{broad} spectral features to the underlying stellar population. No clear evidence is found for broad spectral features associated with the transient at late times.

We propose a model that explains the double-peaked temporal structure of ASASSN-15lh in the optical/UV band as originating from the temporal evolution of the ejecta opacity, which changes as a result of persistent ionizing flux from a long-lived central source (either a magnetar or an accreting SMBH). We speculate that  the evolution of ASASSN-15lh towards a featureless spectrum also results from the presence of a persistent central source of ionizing photons. The exceptionally long active time-scale and high luminosity of the ionizing central source powering ASASSN-15lh (i.e. months) is most likely the key physical property that distinguishes ASASSN-15lh from all the TDEs and SLSNe discovered so far. 

The optical/UV spectral evolution of ASASSN-15lh, its peculiar re-brightening and the presence of soft and persistent X-ray emission are indeed unprecedented among SLSNe and TDEs and suggest two scenarios: (i) either ASASSN-15lh is the first member of a class of stellar explosions with extreme properties that are intrinsically rare or that have been overlooked because of the very close location to the host-galaxy nucleus or, alternatively,  (ii) ASASSN-15lh results from refreshed nuclear activity of the host-galaxy SMBH. 

In the first scenario the detected X-ray emission is physically unrelated to the transient and most likely originates from the host galaxy nucleus. We thus expect no fading of the X-ray source over the time scales of years.

Instead, if the X-ray emission is physically associated with the optical/UV transient, then ASASSN-15lh is unlikely to originate from a stellar explosion and an association with the activity of the host nucleus is favored.  In this case, ASASSN-15lh would be a TDE from the most massive spinning SMBH observed to date.  The fast initial decay timescale of the transient is challenging to understand based on the fall-back timescale of the disrupted star in Newtonian gravity, possibly suggesting that BH spin plays a key role in enhancing the energy spread of the disrupted star.  ASASSN-15lh and similar events discovered in the future would then constitute direct probes of matter under strong gravity around  
very massive, dormant, spinning SMBH in galaxies. We emphasize that this scenario predicts significant temporal evolution of the X-ray emission over the next few years, as we expect a TDE to have a non-negligible impact on the inner part of the accretion disk even in the case of a pre-existing weak AGN.

Continued deep X-ray monitoring of ASASSN-15lh will constrain the temporal evolution of the X-ray source and its fading, revealing in this way if the X-ray source is indeed physically related to the optical/UV transient.  Future X-ray observations thus hold the keys to unveil the true nature of ASASSN-15lh. 
%or would be unique signpost of very massive, dormant, spinning SMBH in galaxies

%%%%%%%%%%%%%%%%%%%%%%%%%%%%%%%%%%%%%%%%%%%
\acknowledgments 
R. M. acknowledges partial support from the James Arthur Fellowship at NYU during the completion of this project and the Research Corporation for Science Advancement. BDM gratefully acknowledges support from the NSF (AST-1410950, AST-1615084), NASA Astrophysics Theory Program (NNX16AB30G), the Alfred P. Sloan Foundation, and the Research Corporation for Science Advancement.   G.M. acknowledges the financial support from the UnivEarthS Labex programof Sorbonne Paris Cite (ANR10LABX0023 and ANR11IDEX000502).
The scientific results reported in this article are based on observations made by the Chandra X-ray Observatory under program GO 17500103, PI Margutti, observations IDs 17879, 17880, 17881, 17882. 
This paper includes data gathered with the 6.5 meter Magellan Telescopes located at Las Campanas Observatory, Chile.
%%%%%%%%%%%%%%%%%%%%%%%%%%%%%%%%%%%%%%%%%%%

\textit{Facilities:} \facility{Magellan:Clay}

\bibliographystyle{apj}
%\bibliography{margutti}

%%%%%%%%%%%%%%%%%%%%%%%%%%%%%%%%%%%%%%%%%%%
\appendix
\section{X-ray and UV/optical Photometry Tables}
\label{App}

\begin{deluxetable}{lllcccccc}
\tabletypesize{\scriptsize}
%\rotate
\tablecolumns{9} 
\tablewidth{0pc}
\tablecaption{X-ray observations. Fluxes are reported in the 0.3-10 keV energy band. We use $NH_{gal}=3.07\times 10^{20}\,\rm{cm^{-2}}$ and a temperature $T=0.17$ keV for the blackbody model (``BB" in the table) and a photon index $\Gamma=3$ for the power-law model (``PL" in the table). Uncertainties are dominated by the choice of the spectral parameters. For Swift-XRT the reported uncertainties reflect the count-rate statistics only.}
\tablehead{ \colhead{Date (MJD)} & \colhead{Instrument} & 
\colhead{Exposure (ks)} & \colhead{Unabsorbed  Flux ($\rm{erg\,s^{-1}cm^{-2}}$)} & \colhead{Spectral Model}}
\startdata
57046-57591  &    Swift/XRT   &  270  & $F_x=(4.1\pm1.5)\times 10^{-15}$ & BB\\
			&				&		&  $F_x=(3.7\pm 1.4)\times 10^{-15}$ & PL\\
57338  &    CXO   &  10   & $F_x<2.0\times 10^{-15}$ & PL \\
		&		&		 & $F_x<8.0\times 10^{-16}$ & BB \\
57344  &    XMM   &   9  &  $F_x<1.5\times 10^{-14}$ & PL \\
  &			&		& $F_x<1.4\times 10^{-14}$ & BB \\
57369  &    CXO   &  10   & $F_x\sim4.4\times 10^{-15}$ & PL \\
		&		&		 & $F_x\sim 1.6\times 10^{-15}$ & BB \\
57438 &    CXO   &  40   & $F_x\sim3.6\times 10^{-15}$ & PL  \\
      &			 & 		& $F_x\sim 1.2\times 10^{-15}$ & BB \\
57619  &    CXO   &  30   & $F_x\sim4.9\times 10^{-15}$ &PL \\
      &			 & 		& $F_x\sim 1.4 \times 10^{-15}$ & BB\\
\enddata
\label{Tab:Xray}
\end{deluxetable}

\begin{deluxetable}{lllcccccccccccc}
\tabletypesize{\scriptsize}
%\rotate
\tablecolumns{12} 
\tablewidth{0pc}
\tablecaption{Swift UVOT photometry}
\tablehead{ \colhead{Date } & \colhead{v (mag)} & 
\colhead{Date } & \colhead{b (mag)} & \colhead{Date }  & \colhead{u (mag)} & \colhead{Date } & \colhead{w1  (mag)} & \colhead{Date} & \colhead{w2 (mag)} & \colhead{Date} & \colhead{m2 (mag)}}
\startdata
197.10\footnote{Dates are in MJD-57000 (days).}  & 16.86(0.07) & 197.09 & 16.76(0.04) & 197.09 & 15.39(0.04) & 197.09 & 15.27(0.04) & 197.10 & 15.63(0.04) & 197.10 & 15.25(0.04) \\ 
199.79 & 16.85(0.07) & 199.79 & 16.82(0.04) & 199.79 & 15.43(0.04) & 199.79 & 15.33(0.04) & 199.79 & 15.67(0.04) & 199.79 & 15.35(0.04) \\ 
201.82 & 16.93(0.08) & 201.82 & 16.87(0.04) & 201.82 & 15.48(0.04) & 201.82 & 15.40(0.04) & 201.82 & 15.83(0.04) & 201.82 & 15.48(0.05) \\ 
205.64 & 16.98(0.08) & 205.64 & 16.94(0.05) & 205.64 & 15.54(0.04) & 205.63 & 15.56(0.04) & 205.64 & 15.88(0.04) & 205.64 & 15.58(0.04) \\ 
208.66 & 17.05(0.08) & 208.66 & 17.03(0.05) & 208.66 & 15.72(0.04) & 208.66 & 15.74(0.04) & 208.66 & 16.07(0.04) & 208.67 & 15.76(0.04) \\ 
211.66 & 17.07(0.09) & 211.66 & 17.05(0.05) & 211.66 & 15.77(0.04) & 211.66 & 15.83(0.05) & 211.66 & 16.18(0.05) & 211.66 & 15.94(0.04) \\ 
221.69 & 17.27(0.13) & 214.68 & 17.09(0.05) & 214.68 & 15.85(0.04) & 214.67 & 15.95(0.05) & 214.68 & 16.35(0.04) & 214.68 & 16.06(0.05) \\ 
223.52 & 17.32(0.10) & 221.68 & 17.29(0.07) & 216.56 & 15.90(0.03) & 219.79 & 16.26(0.06) & 217.01 & 16.48(0.04) & 221.69 & 16.52(0.05) \\ 
231.07 & 17.51(0.08) & 223.51 & 17.27(0.06) & 221.68 & 16.03(0.06) & 221.68 & 16.31(0.06) & 220.69 & 16.71(0.05) & 223.52 & 16.68(0.05) \\ 
244.89 & 17.54(0.11) & 229.27 & 17.41(0.06) & 220.52 & 16.08(0.04) & 223.51 & 16.45(0.05) & 223.51 & 16.90(0.05) & 229.27 & 16.99(0.05) \\ 
255.12 & 17.58(0.07) & 232.89 & 17.45(0.06) & 223.51 & 16.18(0.05) & 229.27 & 16.74(0.06) & 229.27 & 17.25(0.06) & 232.90 & 17.26(0.06) \\ 
268.79 & 17.59(0.12) & 244.88 & 17.71(0.07) & 229.27 & 16.39(0.05) & 232.89 & 16.99(0.06) & 232.89 & 17.50(0.06) & 244.89 & 17.70(0.06) \\ 
310.35 & 17.80(0.08) & 250.70 & 17.49(0.09) & 232.89 & 16.48(0.05) & 244.88 & 17.43(0.07) & 244.88 & 17.99(0.07) & 250.71 & 17.81(0.10) \\ 
383.29 & 18.18(0.32) & 253.70 & 17.66(0.10) & 244.88 & 16.76(0.06) & 250.70 & 17.31(0.10) & 250.71 & 18.07(0.11) & 253.63 & 17.76(0.13) \\ 
214.68 & 17.16(0.09) & 259.40 & 17.87(0.08) & 250.70 & 16.83(0.09) & 253.67 & 17.43(0.08) & 253.70 & 18.11(0.12) & 259.41 & 18.09(0.08) \\ 
226.52 & 17.39(0.07) & 265.56 & 17.82(0.07) & 253.70 & 16.68(0.08) & 259.40 & 17.55(0.08) & 259.40 & 18.46(0.10) & 268.79 & 18.08(0.09) \\ 
241.13 & 17.57(0.09) & 268.79 & 17.89(0.08) & 259.40 & 16.97(0.07) & 265.56 & 17.51(0.07) & 267.22 & 18.29(0.08) & 283.28 & 17.45(0.06) \\ 
248.47 & 17.51(0.09) & 309.34 & 18.06(0.06) & 265.56 & 17.05(0.06) & 268.78 & 17.51(0.07) & 283.88 & 17.82(0.04) & 290.87 & 17.37(0.06) \\ 
262.73 & 17.67(0.10) & 313.24 & 17.95(0.09) & 268.79 & 17.00(0.07) & 282.88 & 17.18(0.06) & 290.88 & 17.71(0.06) & 293.56 & 17.03(0.06) \\ 
272.67 & 17.68(0.10) & 383.28 & 18.59(0.16) & 282.88 & 16.89(0.05) & 283.41 & 17.14(0.06) & 293.57 & 17.56(0.06) & 296.62 & 17.06(0.06) \\ 
277.36 & 17.55(0.07) & 226.52 & 17.35(0.04) & 283.41 & 16.93(0.05) & 290.87 & 17.16(0.06) & 296.63 & 17.50(0.05) & 299.51 & 17.14(0.06) \\ 
284.57 & 17.77(0.07) & 241.12 & 17.58(0.05) & 290.87 & 16.93(0.05) & 293.56 & 17.03(0.06) & 299.52 & 17.51(0.05) & 302.57 & 17.06(0.06) \\ 
298.65 & 17.84(0.11) & 248.47 & 17.60(0.05) & 293.56 & 16.99(0.05) & 296.62 & 16.90(0.05) & 302.57 & 17.55(0.05) & 307.46 & 16.97(0.05) \\ 
305.49 & 17.90(0.10) & 262.73 & 17.88(0.06) & 296.62 & 17.02(0.05) & 299.51 & 16.96(0.06) & 307.45 & 17.28(0.06) & 309.68 & 16.95(0.06) \\ 
317.77 & 17.93(0.14) & 272.67 & 17.83(0.05) & 299.52 & 16.96(0.05) & 302.57 & 17.00(0.05) & 309.68 & 17.30(0.07) & 311.31 & 16.81(0.05) \\ 
330.19 & 17.93(0.11) & 277.36 & 17.72(0.05) & 302.57 & 16.90(0.05) & 307.45 & 16.93(0.06) & 311.31 & 17.20(0.06) & 313.24 & 16.93(0.05) \\ 
346.20 & 17.92(0.10) & 284.57 & 17.86(0.05) & 307.45 & 16.91(0.06) & 309.67 & 16.81(0.07) & 313.24 & 17.27(0.06) & 353.56 & 16.96(0.07) \\ 
371.42 & 18.19(0.17) & 298.64 & 18.09(0.07) & 309.68 & 16.84(0.08) & 311.30 & 16.85(0.06) & 353.63 & 17.30(0.10) & 356.22 & 17.12(0.07) \\ 
393.66 & 18.10(0.15) & 255.69 & 17.74(0.05) & 311.30 & 16.92(0.07) & 313.23 & 16.83(0.06) & 356.23 & 17.21(0.05) & 359.31 & 17.04(0.07) \\ 
447.41 & 18.49(0.29) & 305.49 & 17.98(0.06) & 313.23 & 16.93(0.07) & 353.56 & 16.91(0.07) & 359.32 & 17.29(0.06) & 362.84 & 17.21(0.08) \\ 
398.03 & 18.02(0.11) & 317.77 & 17.91(0.07) & 353.57 & 16.94(0.09) & 356.23 & 16.96(0.06) & 362.84 & 17.23(0.06) & 365.09 & 17.19(0.07) \\ 
432.60 & 18.59(0.19) & 330.18 & 18.08(0.08) & 356.23 & 17.01(0.06) & 359.32 & 17.05(0.06) & 365.73 & 17.26(0.06) & 374.40 & 17.14(0.07) \\ 
452.20 & 18.33(0.17) & 346.19 & 18.10(0.06) & 359.32 & 16.89(0.06) & 362.84 & 16.96(0.07) & 374.41 & 17.51(0.06) & 377.19 & 17.23(0.07) \\ 
456.92 & 18.24(0.17) & 366.43 & 18.15(0.11) & 362.84 & 17.22(0.08) & 365.10 & 17.13(0.07) & 377.20 & 17.42(0.05) & 380.91 & 17.16(0.07) \\ 
472.23 & 18.35(0.39) & 371.41 & 18.36(0.10) & 365.10 & 17.16(0.07) & 368.95 & 16.99(0.07) & 380.92 & 17.53(0.06) & 384.52 & 17.31(0.09) \\ 
477.55 & 18.22(0.26) & 393.66 & 18.35(0.11) & 368.95 & 17.04(0.07) & 374.40 & 17.17(0.07) & 384.51 & 17.35(0.06) & 226.52 & 16.89(0.05) \\ 
533.61 & 18.74(0.51) & 427.75 & 18.64(0.13) & 374.41 & 17.23(0.07) & 377.19 & 17.29(0.08) & 226.52 & 17.16(0.05) & 241.13 & 17.61(0.05) \\ 
552.75 & 18.26(0.25) & 447.41 & 19.10(0.22) & 377.20 & 17.20(0.08) & 380.92 & 17.25(0.07) & 241.12 & 17.77(0.06) & 248.48 & 17.83(0.06) \\ 
554.88 & 18.02(0.17) & 396.75 & 18.56(0.13) & 380.92 & 17.22(0.08) & 384.51 & 17.18(0.08) & 248.47 & 18.12(0.06) & 262.73 & 18.16(0.07) \\ 
557.37 & 18.52(0.17) & 399.25 & 18.62(0.14) & 384.51 & 17.22(0.09) & 226.52 & 16.62(0.05) & 262.73 & 18.38(0.07) & 272.68 & 17.88(0.07) \\ 
560.22 & 18.57(0.19) & 433.43 & 18.86(0.10) & 226.52 & 16.30(0.04) & 241.12 & 17.25(0.05) & 272.67 & 18.22(0.06) & 277.33 & 17.62(0.07) \\ 
563.21 & 18.82(0.40) & 452.19 & 18.91(0.14) & 241.12 & 16.63(0.05) & 248.47 & 17.43(0.06) & 277.36 & 17.87(0.05) & 284.57 & 17.32(0.05) \\ 
569.44 & 18.42(0.27) & 456.91 & 18.91(0.13) & 248.47 & 16.78(0.05) & 262.73 & 17.55(0.06) & 298.64 & 17.54(0.05) & 298.65 & 17.11(0.05) \\ 
 &  & 472.22 & 19.14(0.30) & 262.73 & 17.08(0.06) & 272.67 & 17.39(0.05) & 255.70 & 18.29(0.07) & 255.66 & 18.02(0.09) \\ 
 &  & 477.55 & 19.03(0.21) & 272.88 & 17.01(0.07) & 277.35 & 17.15(0.05) & 305.49 & 17.34(0.05) & 305.50 & 17.03(0.06) \\ 
 &  & 533.61 & 18.79(0.23) & 277.36 & 16.95(0.05) & 284.37 & 17.11(0.05) & 317.77 & 17.39(0.06) & 317.77 & 16.83(0.09) \\ 
 &  & 552.75 & 19.01(0.19) & 284.37 & 16.97(0.05) & 298.64 & 17.01(0.05) & 330.19 & 17.28(0.05) & 330.19 & 16.97(0.05) \\ 
 & & 554.88 & 19.13(0.16) & 298.64 & 16.88(0.05) & 255.69 & 17.51(0.06) & 346.20 & 17.35(0.05) & 346.20 & 16.98(0.06) \\ 
 & & 557.36 & 19.09(0.12) & 255.69 & 16.91(0.05) & 284.77 & 17.12(0.06) & 371.42 & 17.41(0.05) & 371.43 & 17.25(0.07) \\ 
 & & 560.22 & 19.12(0.13) & 272.60 & 16.96(0.05) & 305.48 & 16.87(0.05) & 316.24 & 17.27(0.05) & 316.23 & 16.98(0.07) \\ 
 & & 563.21 & 19.04(0.21) & 284.77 & 16.95(0.05) & 317.73 & 17.02(0.06) & 319.77 & 17.43(0.06) & 319.76 & 17.13(0.07) \\ 
 & & 569.43 & 19.41(0.24) & 305.49 & 16.87(0.05) & 330.18 & 16.92(0.05) & 325.34 & 17.52(0.06) & 322.75 & 17.19(0.07) \\ 
 & &  & & 317.77 & 16.94(0.06) & 346.19 & 17.02(0.05) & 328.53 & 17.30(0.05) & 325.33 & 17.22(0.07) \\ 
 & &  & & 330.18 & 17.00(0.06) & 366.39 & 17.08(0.07) & 331.06 & 17.24(0.05) & 328.52 & 16.96(0.06) \\ 
 & &  & & 346.19 & 16.97(0.05) & 371.41 & 17.16(0.06) & 334.58 & 17.32(0.06) & 331.05 & 17.03(0.07) \\ 
 & &  & & 366.43 & 17.18(0.08) & 316.23 & 16.95(0.06) & 343.44 & 17.34(0.06) & 334.58 & 17.02(0.07) \\ 
 & &  & & 371.41 & 17.17(0.07) & 321.23 & 16.98(0.06) & 349.22 & 17.27(0.06) & 343.43 & 17.06(0.06) \\ 
 & &  & & 316.24 & 16.87(0.06) & 325.34 & 16.98(0.06) & 393.66 & 17.55(0.05) & 349.21 & 16.94(0.07) \\ 
 & &  & & 319.76 & 17.01(0.06) & 328.53 & 16.78(0.06) & 427.75 & 18.16(0.07) & 393.67 & 17.37(0.07) \\ 
 & &  & & 325.34 & 16.92(0.06) & 331.05 & 16.82(0.06) & 447.41 & 18.06(0.07) & 427.75 & 17.83(0.08) \\ 
 & &  & & 328.53 & 16.86(0.05) & 334.58 & 16.80(0.06) & 396.76 & 17.55(0.06) & 447.41 & 17.87(0.08) \\ 
 & &  & & 331.05 & 16.86(0.06) & 343.43 & 16.93(0.06) & 399.25 & 17.41(0.06) & 396.76 & 17.31(0.07) \\ 
 & &  & & 334.58 & 16.97(0.06) & 349.21 & 16.88(0.06) & 429.65 & 18.06(0.09) & 399.25 & 17.29(0.07) \\ 
 & &  & & 343.43 & 17.03(0.06) & 393.66 & 17.27(0.06) & 433.51 & 18.15(0.09) & 429.65 & 17.83(0.10) \\ 
 & &  & & 349.21 & 16.98(0.06) & 427.74 & 17.59(0.07) & 437.32 & 18.16(0.08) & 435.30 & 17.84(0.08) \\ 
 & &  & & 393.66 & 17.37(0.07) & 447.41 & 17.73(0.09) & 450.04 & 18.09(0.08) & 450.04 & 18.00(0.09) \\ 
 & &  & & 427.75 & 17.73(0.09) & 396.75 & 17.20(0.07) & 454.22 & 18.17(0.08) & 454.22 & 17.90(0.09) \\ 
 & &  & & 447.41 & 18.00(0.12) & 399.25 & 17.27(0.07) & 457.64 & 18.19(0.10) & 456.09 & 17.89(0.09) \\ 
 & &  & & 396.75 & 17.36(0.08) & 429.64 & 17.74(0.10) & 456.08 & 18.03(0.07) & 465.55 & 18.01(0.08) \\ 
 & &  & & 399.25 & 17.33(0.07) & 433.50 & 17.76(0.10) & 465.56 & 18.17(0.07) & 470.33 & 17.92(0.07) \\ 
 & &  & & 431.64 & 17.85(0.10) & 437.32 & 17.71(0.09) & 470.34 & 18.19(0.06) & 472.23 & 18.05(0.16) \\ 
 & &  & & 437.32 & 17.75(0.11) & 450.03 & 17.82(0.12) & 472.23 & 18.18(0.13) & 475.67 & 17.94(0.08) \\ 
 & &  & & 450.03 & 17.85(0.13) & 454.21 & 17.78(0.09) & 475.68 & 18.29(0.07) & 477.55 & 18.06(0.11) \\ 
 & &  & & 454.21 & 17.80(0.11) & 457.64 & 17.92(0.12) & 477.55 & 18.26(0.10) & 480.30 & 18.20(0.09) \\ 
 & &  & & 456.91 & 17.81(0.08) & 456.07 & 17.83(0.08) & 480.30 & 18.39(0.07) & 487.48 & 18.29(0.15) \\ 
\enddata
\label{Tab:UVOT}
\end{deluxetable}

\end{document}